# Efficient Collision Algorithms in DSMC for Rarefied Gas Dynamics: Markovian NTC-Pre-Scan and Bernoulli-Trial Schemes

Ahmad Shoja-sani[1], Ehsan Roohi[2*], Maryam Javani[1], Hassan Akhlaghi[3], Stefan Stefanov[4, 5]

[1] High Performance Computing (HPC) Laboratory, Department of Mechanical Engineering, Ferdowsi University of Mashhad, 91775-1111, Mashhad, Iran
[2] Mechanical and Industrial Engineering, University of Massachusetts Amherst, 160 Governors Dr., Amherst, MA 01003, USA. (*Corresponding author: roohie@umass.edu)
[3] Department of Aerospace Engineering, College of Interdisciplinary Science and Technologies, University of Tehran.Tehran, Iran
[4] Institute of Mechanics, Bulgarian Academy of Science, Acad. G. Bontchev Str., 1113 Sofia, Bulgaria
[5] Centre of Excellence in Informatics and Information and Communication Technologies, Sofia, Bulgaria

**Abstract**

The collision process is essential to the Direct Simulation Monte Carlo (DSMC) method, as it incorporates the fundamental principles of the Boltzmann and Kac stochastic equations. A series of collision algorithms, known as the Bernoulli-trials (BT) family schemes, have been proposed based on the Kac stochastic equation. The primary impetus of this paper is to rectify a long-standing theoretical flaw in the widely used no-time-counter (NTC) collision algorithm. We demonstrate that the standard NTC scheme is fundamentally non-Markovian, relying on a fixed majorant product that introduces a system 'memory' and leads to inaccuracies at low particle counts. We propose a new algorithm, NTC-Pre-Scan, which transforms the scheme into a fully Markovian process. When the repeated collisions are not crucial, our new NTC scheme, called NTC-Pre-Scan, could work accurately with a very low number of particles per cell (PPC), average PPC<1, i.e., PPC=0.01, which means with several empty cells in simulations. This contrasts with the standard NTC schemes, which typically require a PPC greater than 1. Then, a systematic evaluation of different BT-based collision partner selection schemes, including the simplified Bernoulli trials (SBT), generalized Bernoulli trials (GBT), symmetrized and simplified Bernoulli trials (SSBT), and the newly proposed symmetrized and generalized Bernoulli trials (SGBT), is conducted to treat some benchmark rarefied gas dynamics problems. The results show that the BT-based collision algorithms and NTC-Pre-scan successfully maintain the collision frequency as the number of particles per cell decreases. Simulation of the Bobylev-Krook-Wu (BKW), where an exact solution of the Boltzmann equation is available, indicates that, like the GBT, the SGBT algorithm produces the same results as the theory for the average of the fourth moment of the velocity distribution function (VDF). The simulation on the three-dimensional computational grid for the GBT and SGBT schemes matches the fourth moment of the velocity component of the VDF exactly with the analytical solution. Performance analysis in a micro cavity reveals that the GBT, SSBT, and SGBT decrease the computational cost of simulation. Specifically, the computational cost of the SGBT scheme has been reduced by around 40% when an appropriate selection number ($N_{sel}$) is chosen, and this scheme requires a sample size of 0.62 of the NTC scheme. Finally, we demonstrate that all algorithms successfully capture complex flow phenomena, such as shock waves, in the case of hypersonic flow over a cylinder. Moreover, in the cylinder problem, the SGBT scheme can achieve the same level of accuracy with 28% less computational cost and an outstanding sample size of 0.319 of the nearest neighbor (NN) scheme, which is the modern invariant of the NTC scheme. These advancements enable accurate simulation of rarefied gases with fewer particles (NTC-Pre-Scan) and lower computational cost (SGBT), which is directly beneficial for the design of complex vacuum systems.

## 1. Introduction

The Boltzmann equation provides a fundamental description of fluid dynamics across a wide range of Knudsen numbers, from the continuum to the free-molecular regimes. A cornerstone for its solution is the Direct Simulation Monte Carlo (DSMC) method [1-2], which probabilistically models particle kinetics and asymptotically converges to the solution of the Boltzmann equation. Different approaches to solving the Boltzmann equation, including the Fokker-Planck model [3] and the hybrid DSMC-NS approach [4], were reported in the literature. DSMC has been widely used for simulating practical vacuum-based pumps and devices, see for example Ref. [5]. Recognizing the strengths of different approaches, researchers have also introduced innovative multiscale methods that intermittently couple DSMC with deterministic solvers to enhance accuracy and accelerate convergence on coarse grids [6]. More recently, a new frontier has emerged with the application of machine learning to rarefied gas dynamics [7–10].

The simulation of rarefied gas flows in vacuum systems—such as leak detection in vacuum chambers, the design of microchannels in micro/nano-electro-mechanical systems (MEMS/NEMS), or the optimization of molecular pumps—is commonly performed using the DSMC method. In these applications, both the speed and accuracy of DSMC are critical, as they directly determine the feasibility and reliability of vacuum system design and analysis. Given its foundational role both as a benchmark solver and as a generator of high-fidelity data for these emerging techniques, it is pertinent to detail the principles of the DSMC method. As a widely recognized and extensively utilized computational technique for modeling rarefied gas flows, particularly in non-equilibrium conditions, DSMC was initially introduced by Graeme Bird [1]. The method simulates gas behavior by tracking the motion and collisions of representative model particles (simulators) within physical space. Unlike Molecular Dynamics (MD), which resolves individual particle interactions deterministically, DSMC employs a statistical framework to model binary intermolecular collisions, thereby enabling efficient computation of gas flows at the molecular scale.

Using a finite number of simulators, DSMC facilitates the molecular-level modeling of complete flow fields. Comparative studies with experimental data and simplified analytical solutions of the Boltzmann equation (BE) [11-12] demonstrated a strong correlation between DSMC results and the BE. Wagner [13] formally established this connection, providing a rigorous mathematical proof showing the asymptotic convergence of the DSMC method to the Boltzmann equation, thereby positioning DSMC as a practical numerical tool for solving the BE. However, a more rigorous theoretical analysis conducted by Stefanov [14] revealed fundamental subtleties in this interpretation. His investigation showed that the Boltzmann equation does not universally govern the DSMC method. Instead, he derived a new kinetic equation that more accurately describes the underlying dynamics of DSMC, thereby redefining the theoretical foundation of the method.

The DSMC method models molecular interactions by dividing the simulation process into two distinct stages: free molecular motion and intermolecular collisions, both of which occur within discrete computational cells. This structured approach significantly improves the efficiency and fidelity of particle-based gas flow simulations. Among these stages, the collision process is particularly critical, prompting the development of various algorithms designed to accurately capture molecular interactions.

One of the most widely implemented collision algorithms is the No Time Counter (NTC) method, introduced by Bird [15]. This scheme is grounded in the statistical framework of the Boltzmann equation, or more precisely, in the kinetic equation proposed by Stefanov [14]. The NTC method is widely regarded as the gold standard for collision modeling in DSMC and has been highly successful. Nevertheless, there remains motivation for further improvement, particularly in regimes where the number of particles per cell becomes very small, i.e., high-vacuum conditions or very fine grids, or when large-scale or real-time simulations are required. Although improving such a well-established method may seem challenging, there remains interest in refining it for modeling complex vacuum systems. Basic improvement in the NTC scheme, by making it Markovian, will be addressed in this paper.

Another notable approach is the Majorant Frequency Scheme, developed by Ivanov et al. [16], which has also gained widespread use. Despite its popularity, the NTC method has a known limitation: in cells containing a small number of particles, it may result in repeated collisions between the same particle pairs. Stefanov [17] demonstrated that a sequence of $n$ consecutive collisions between the same two particles produces a post-collision velocity distribution statistically identical to that of a single collision. This finding implies that multiple successive collisions do not contribute additional physical effects but instead artificially reduce the effective collision rate. Therefore, minimizing repeated collisions is crucial for ensuring the accuracy of the DSMC method, particularly in low-density regimes where the particle count per cell is limited [18].

An alternative class of collision algorithms was developed based on the Kac stochastic equation, which laid the theoretical foundation for particle-based collision modeling. From this formulation, a distinct family of algorithms known as Bernoulli trial (BT) methods was derived by Belotserkovskii and Yanitskiy [19], and further refined by Yanitskiy [20]. The BT scheme was specifically designed to mitigate the issue of repeated collisions inherent in the No Time Counter (NTC) method. Although the BT approach effectively eliminates redundant collisions, it introduces a significant computational burden, as its algorithmic complexity scales with the square of the number of particles in a cell, i.e., $O(N^{(l)})^2$ where $N^{(l)}$ denotes the instantaneous number of particles in cell $l$.

Stefanov [17, 21] proposed a simplified version of the Bernoulli Trial (BT) algorithm, known as the Simplified Bernoulli Trial (SBT) scheme. The performance of the SBT method has been comprehensively assessed in several studies [22–24], demonstrating its effectiveness in reducing repeated collisions while maintaining computational efficiency. To further improve the efficiency of the SBT approach, Roohi et al. [25] developed the Generalized Bernoulli Trial (GBT) algorithm, which reduced computational overhead by optimizing the selection process during collision handling. Subsequently, Taheri et al. [26] introduced the Symmetrized and Simplified Bernoulli Trial (SSBT) scheme, an enhanced version of BT that accounts for all particles before and after the selected particle within a cell. This symmetric consideration offers a more physically realistic representation of molecular collisions, thereby improving the accuracy of simulating gas behavior. Most recently, Javani et al. [27] proposed a hybrid collision scheme, SGBT, that integrates key features of both the GBT and SSBT models. Like SSBT, this method evaluates potential collisions with all neighboring particles while adopting the $N_{sel}$-based selection strategy from GBT to lower computational costs. This combined approach effectively balances accuracy and efficiency, making it suitable for high-fidelity DSMC simulations.

Although the post-collision calculation (the randomization of molecular velocities after collisions) is identical in all these methods and does not introduce additional complexity, the collision-pair selection step accounts for a significant portion of the computational cost in DSMC. In other words, improving this step directly accelerates the entire simulation without making the post-collision stage more complicated. The results provided later in this paper demonstrate that even while preserving the simplicity of the post-collision calculation, a more efficient choice of collision pairs can lead to a higher overall simulation speed. To summarize the collision schemes introduced so far, a general schematic of these collision schemes is briefly shown in Figure 1. The NTC collision scheme selects the particle pair from all the particles in the cell. In the following, the collision schemes are based on the Kac equation, where the black particle represents the first collision particle, and the remaining particles, shown in white, can be selected as the collision pair of the black particle. All potential pairs are evaluated in the BT algorithm, as illustrated by the arrows extending from each black circle. In contrast, the SBT algorithm connects each arrow to a set of particles from which a second partner is randomly selected. The GBT algorithm consists of two steps, both of which are illustrated in the selection process. Initially, a specified number of particles, denoted as $N_{sel}$, is selected from the entire set within the cell or subcell. In the subsequent step, the SBT method is applied to the $N_{sel}$ particles represented by the black circles. In the BB procedure, a single collision pair is selected. Both the black circle representing the first selected particle and the second particle are chosen at random. This algorithm identifies collision pairs by integrating the SSBT and GBT schemes. In the initial step, like SSBT, each particle has the potential to collide with any particle present in the list, whether before or after the selected particle.

In the subsequent step, mirroring the GBT approach, the selected pairs can take on any arbitrary value less than $N^l$, such that $1 < N_{sel} < N^l$.

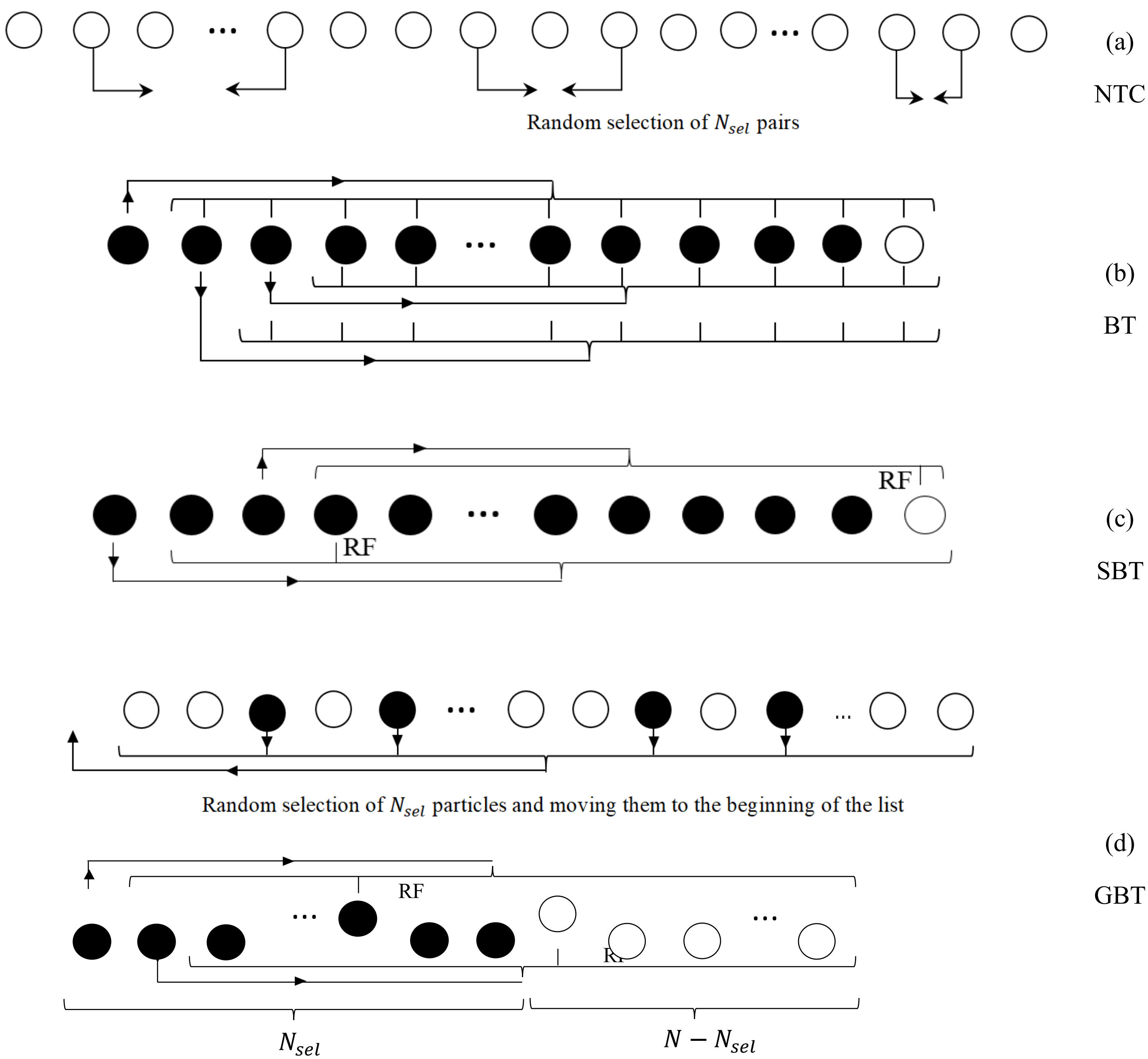

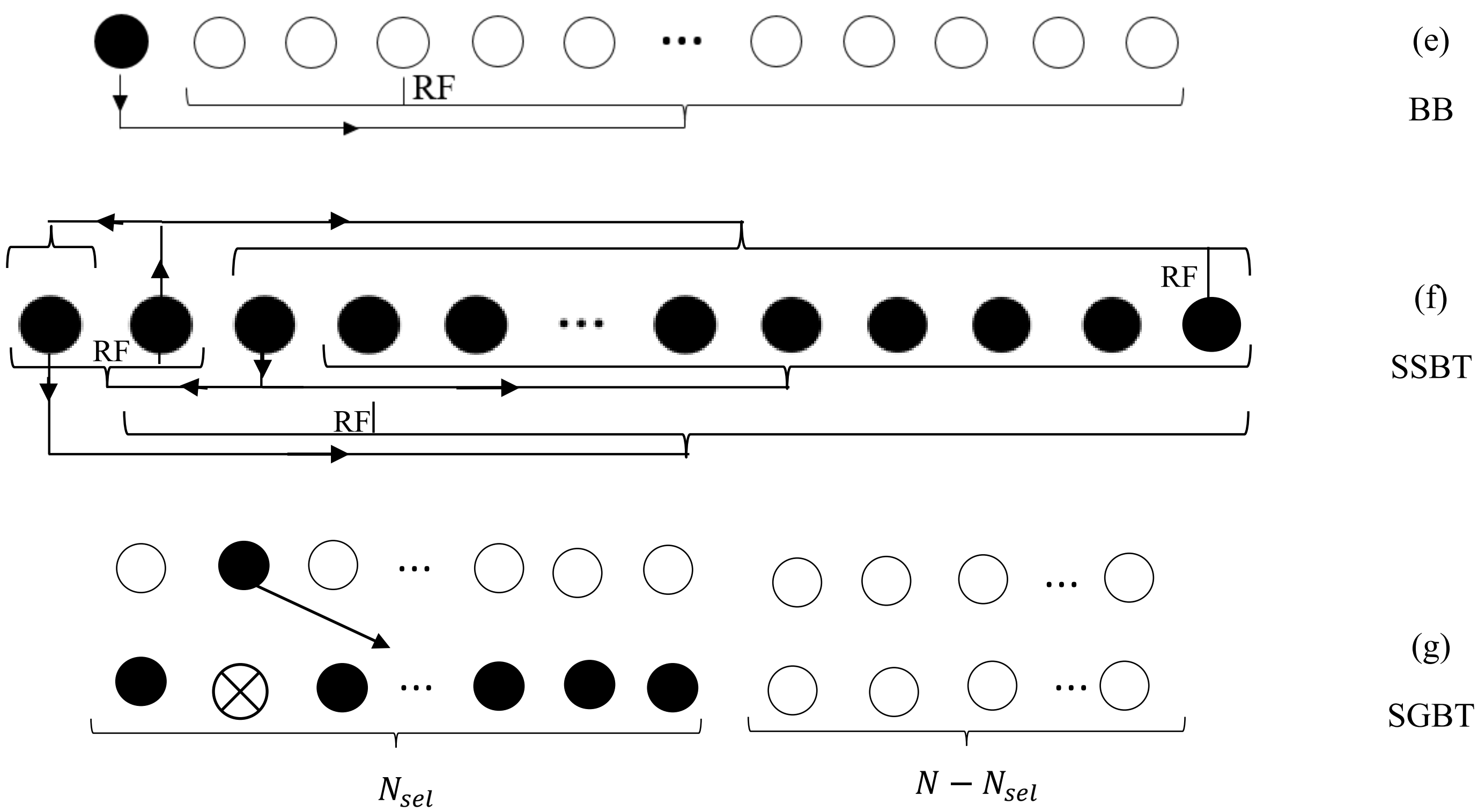

Fig 1. Schematic picture of various collision models, NTC, BT, SBT, GBT, BB, SSBT, SGBT. NTC selects randomly from all pairs, BT selects all particles in the list after the first chosen one consecutively from the list, SBT selects one particle from the list after the first chosen particle, GBT follows the SBT strategy but limits it to $N_{sel}$ particles, not all the particles in the list, BB selects just one particle from the list of particles after the first one (checks one pair), SSBT is like SBT but selects the second particle from everywhere in the list, and SGBT follows SSBT algorithm for $N_{sel}$ particles. RF stands for random number.

In this study, we present a comprehensive assessment of the aforementioned collision partner selection schemes by applying them to several fundamental problems in rarefied gas flow. The primary goal of this paper is to correct a long-standing theoretical flaw in the widely used NTC collision algorithm. We show that the standard NTC scheme is inherently non-Markovian, relying on a fixed majorant product that introduces a system 'memory' and causes inaccuracies at low particle counts. We introduce a new algorithm, NTC-Pre-Scan, which makes the scheme fully Markovian. When repeated collisions are not critical, our new NTC scheme, called NTC-Pre-Scan, can operate accurately with a very low number of particles per cell (PPC), averaging less than 1

PPC, for example, PPC=0.01, meaning several cells may be empty in simulations. Second, we investigate a series of test cases to show that the BT family scheme can achieve the same level of accuracy as NTC but with a suitably reduced computational time and a smaller sample size. The selected test cases include: a collision frequency benchmark, the spatially homogeneous relaxation of a gas from a non-Maxwellian initial condition based on the Bobylev–Krook–Wu (BKW) analytical solution of the Boltzmann equation, lid-driven microcavity flow, and hypersonic flow over a circular cylinder. Importantly, the problems studied here were selected not because they are generic test cases, but because they embody diverse physical regimes—ranging from homogeneous relaxation to multidimensional cavity and shock-dominated flows—allowing us to reveal nontrivial characteristics of each algorithm under different rarefaction levels. They involve non-isothermal walls, strong gradients, and near-wall phenomena that are highly sensitive to collision modeling. Thus, our work offers not only performance benchmarking but also fundamental insights into how collision algorithms influence rarefied gas dynamics in both equilibrium and non-equilibrium regimes.

For the zero-dimensional (0-D) test cases, all collision schemes are implemented within Bird's DSMC0F code. In addition, an advanced Python code was developed for this problem, which reports detailed statistics of the flow field from various collision schemes. To simulate the microcavity flow, the DSMC2 code by Bird is modified to incorporate the selected collision models. The more complex two-dimensional (2D) problem of hypersonic flow over a circular cylinder is addressed using Bird's advanced DSMC solver, DS2V, which is also adapted to accommodate the proposed algorithms. Our findings consistently show that the proposed SGBT scheme maintains high physical fidelity and efficiency across all problem types and Knudsen numbers, while reducing the sample size and computational costs with an appropriate number of selections. This reduction is achieved because this scheme selects and tests fewer particle pairs to achieve an accurate solution.

## 2. Bernoulli Trials Collision Methods

Stefanov [14] formulated a new nonhomogeneous, local kinetic equation that describes the evolution of the DSMC method in terms of an *N*-particle system. This equation is derived from the discretized form of the kinetic equations, obtained through operator splitting in both time and space (see Eqs. (2)–(4)), and is applied to the *N*-particle distribution function ($\overline{F}_N$). The formulation accounts for a random number of particles with an expected value $s = \overline{N}^{(l)}$, and considers the set of particle velocities $\mathbb{C}=\{c_1, \ldots, c_i, \ldots, c_j, \ldots, c_N\}$ observed at time t within a small control volume *V* centered around a spatial location **r.** This equation has the following form:

$$\frac{\partial \bar{F}_N(t, \boldsymbol{r}, \mathbb{C})}{\partial t} + \sum_{i=1}^{N} \boldsymbol{c}_i \frac{\partial \bar{F}_N(t, \boldsymbol{r}, \mathbb{C})}{\partial \boldsymbol{r}} = \sum_{1 \le i < j \le N} \left\{ \int g_{i,j} \left[ \bar{F}_N\left(t, \boldsymbol{r}, \mathbb{C}'_{i,j}\right) - \bar{F}_N(t, \boldsymbol{r}, \mathbb{C}) \right] d\sigma_{i,j} \right\} \quad (1)$$

, where $\mathbb{C}'_{i,j}=\{c_1, \ldots, c'_i, \ldots, c'_j, \ldots, c_N\}$ is the velocity vector after collision between particles *i* and *j*, $g_{i,j}=|c_i-c_j|$ and $d\sigma_{j,j}$ is the differential collision cross-section. This equation is defined in three-dimensional physical space and 3*N*-dimensional velocity space, describing the evolution of the local *N*-particle distribution in the vicinity of the physical point r ∈ $R^3$.

Within a small discrete time interval $\Delta t$ the splitting form of equations is presented [14] as follows:

$$t < \tau \le t + \Delta t, \; l = 1, M \quad (2)$$

$$\left|\begin{array}{l}\tilde{F}^{*}{}_{N^{(l)}}\left(t+0,\boldsymbol{r}^{(l)},\mathbb{C}^{(l)}\right)=\tilde{F}_{N^{(l)}}\left(t,\boldsymbol{r},\mathbb{C}^{(l)}\right),\quad \boldsymbol{r}\in D^{(l)}\subset R^{3}\\ \dfrac{\partial\tilde{F}^{*}{}_{N^{(l)}}\left(t,\boldsymbol{r}^{(l)},\mathbb{C}^{(l)}\right)}{\partial t}=\dfrac{1}{V^{(l)}}\sum_{1\leq i<j\leq N^{(l)}}\left\{\int g_{i,j}\left[\tilde{F}^{*}{}_{N^{(l)}}\left(t,\boldsymbol{r}^{(l)},\mathbb{C}'^{(l)}_{i,j}\right)-\tilde{F}^{*}{}_{N^{(l)}}\left(t,\boldsymbol{r}^{(l)},\mathbb{C}^{(l)}\right)\right]d\sigma_{i,j}\right\}\end{array}\right.$$

$$\left|\begin{array}{l}\tilde{F}^{**}{}_{N^{(l)}}\left(t+0,\boldsymbol{r}^{(l)},\mathbb{C}^{(l)}\right)=\tilde{F}^{*}{}_{N^{(l)}}\left(t+\Delta t,\boldsymbol{r}^{(l)},\mathbb{C}^{(l)}\right),\\ \dfrac{\partial\tilde{F}^{**}{}_{N^{(l)}}\left(t,\boldsymbol{r}^{(l)},\mathbb{C}^{(l)}\right)}{\partial t}=-\sum_{i=1}^{N^{(l)}}\boldsymbol{c}_{i}\dfrac{\partial\tilde{F}^{**}{}_{N^{(l)}}\left(t,\boldsymbol{r}^{(l)},\mathbb{C}^{(l)}\right)}{\partial\boldsymbol{r}}\qquad \boldsymbol{r}\in D^{(l)}\end{array}\right. \tag{3}$$

$$\left|\begin{array}{l}F_{N}(t+\Delta t,\boldsymbol{r},\mathbb{C})=\sum_{l=1}^{M}\tilde{F}^{**}{}_{N^{(l)}}\left(t+\Delta t,\boldsymbol{r}^{(l)},\mathbb{C}^{(l)}\right)\\ \tilde{F}_{N^{(l)}}\left(t+\Delta t,\boldsymbol{r}^{(l)},\mathbb{C}^{(l)}\right)=\int_{D^{(l)}}F_{N}(t+\Delta t,\boldsymbol{r},\mathbb{C})d\boldsymbol{r},\quad \boldsymbol{r}\in D\subset R^{3}\end{array}\right. \tag{4}$$

, where Δt is the time step and *M* represents the number of subdomain $D^{(l)}$ in the spatial domain *D* of $R^3$. In this manner, $D^{(l)}$ is analogous to a computational grid cell in the DSMC. $\tilde{F}_N$ is the *N*-particle distribution function of particle velocities in the vicinity of a spatial point $r\in D\subset R^3$. Equation (2), which describes the evolution of the collision process of the stochastic model in a time interval $\tau\in(t,t+\Delta t)$ can be used to derive different collision schemes. Yanitskiy [20,28] has proposed the following compact operator solution for this equation:

$$\tilde{F}^{*}{}_{N^{(l)}}\left(t+\Delta t,\boldsymbol{r}^{(l)},\mathbb{C}^{(l)}\right)=G(\Delta t)\tilde{F}^{*}{}_{N^{(l)}}\left(t,\boldsymbol{r}^{(l)},\mathbb{C}^{(l)}\right), \tag{5}$$

where the transition operator $G(\Delta t)$ is given by

$$G(\Delta t)=\sum_{k=0}^{\infty}\frac{\Omega^{k}}{k!}(\nu\Delta t)^{k}=exp[\Delta t\nu(T-I)] \tag{6}$$

The new operators $I$and $T$, introduced in (6), decompose operator $\Omega$ into identity and 3D-velocity rotation operators

$$I\,\psi=\psi,\quad T\,\psi=\frac{1}{\sigma_{i,j}}\int_{4\pi}\psi\left(\boldsymbol{c}_{i},\boldsymbol{c}_{j}\right)\sigma\left(g_{i,j},\theta\right)d\theta d\varepsilon\,,T\,\psi=\sum_{1\leq i<j\leq N^{(l)}}\omega_{i,j}\,T_{i,j}\,\psi. \tag{7}$$

The unknown binary collision frequency$\nu$is given by

$$\nu=\sum_{1\leq i<j\leq N^{(l)}}\omega_{i,j}\,;\qquad \omega_{i,j}=\frac{\sigma_{i,j}g_{i,j}}{V^{(l)}} \tag{8}$$

Thus, the final form of $G(\Delta t)$could be written as

$$G(\Delta t) = \exp\left[\Delta t \sum_{1 \le i < j \le N^{(l)}} \omega_{i,j} \left(T_{i,j} - I\right)\right] \tag{9}$$

Yanitskiy [28] has shown that the stochastic interpretation of the transition operator (9) gives a general scheme of the collision process in each cell $(l)$ with a collision probability of any randomly chosen particle pair $(i, j)$ in $(l)$equal to

$$W_{i,j} = \frac{\omega_{i,j}}{\nu} = \frac{\omega_{i,j}}{\sum_{1 \le i < j \le N^{(l)}} \omega_{i,j}} \tag{10}$$

The operator (9) can be written in the following form

$$G(\Delta t) = \prod_{i=1}^{N^{(l)}-1} \prod_{j=i+1}^{N^{(l)}} \exp\left[\Delta t \omega_{i,j} \left(T_{i,j} - I\right)\right], \tag{11}$$

and each co-factor in this presentation is replaced by an approximation linear in $\Delta t$. This gives

$$G_{BT}(\Delta t) = \prod_{i=1}^{N^{(l)}-1} \prod_{j=i+1}^{N^{(l)}} \left[\left(1 - \Delta t \omega_{i,j}\right) I + \Delta t \omega_{i,j} T_{i,j}\right] \tag{12}$$

Stefanov [17] showed that it is possible to extend the internal product in (12) in a series of $j$ with respect to $\Delta t$ in order to obtain a new simplified transition operator

$$G_{SBT}(\Delta t) = \prod_{i=1}^{N^{(l)}-1} \left\{\left[1 - \sum_{j=1}^{N^{(l)}} \frac{1}{k}\left(k\omega_{i,j}\Delta t\right)\right] I + \sum_{j=1}^{N^{(l)}} \frac{1}{k}\left(k\omega_{i,j}\Delta t\right) T_{i,j}\right\}, \tag{13}$$

where $k = N^{(l)} - i$. The algorithmic interpretation of the operator $G_{SBT}(\Delta t)$determines the Simplified Bernoulli trials (SBT) scheme. The number of selected pairs per time step is equal to $N^{(l)} - 1$ with a requirement for eventual overruns of condition$\left(k\omega_{i,j}\Delta t\right) \le 1$ to be very rare.

An additional partial linearization of operator (13) about $\Delta t$, which is proposed in the Reference [25], gives a generalized form of the approximation operator

$$G_{GBT}(\Delta t) = \prod_{i=1}^{N_{sel}} \left\{\left[1 - \sum_{j=1}^{N^{(l)}} \frac{1}{k'k}\left(k'k\omega_{i,j}\Delta t\right)\right] I + \sum_{j=1}^{N^{(l)}} \frac{1}{k'k}\left(k'k\omega_{i,j}\Delta t\right) T_{i,j}\right\}, \tag{14}$$

Where the number of selected pair $N_{sel}$ can vary in the range$1 \le N_{sel} \le N^{(l)} - 1$ and $k'k$ is equal to

$$k'k = \frac{N^{(l)}\left(N^{(l)} - 1\right)}{N_{sel}\left(2N^{(l)} - N_{sel} - 1\right)} \left(N^{(l)} - i\right), \tag{15}$$

Next, the symmetric form used to obtain the SSBT generator is as follows [26]:

$$G_{SSBT}(\Delta t) = \prod_{i=1}^{N^{(l)}} \left\{ \left[ 1 - \sum_{j=1, i \neq j}^{N^{(l)}} \frac{1}{k'} \left[ \frac{k'}{2} \left( p_{i,j} + p_{j,i} \right) \right] \right] \boldsymbol{I} + \sum_{j=1, i \neq j}^{N^{(l)}} \frac{1}{k'} \left[ \frac{k'}{2} \left( p_{i,j} + p_{j,i} \right) \right] \boldsymbol{T}_{i,j} \right\}. \quad (16)$$

The internal summation for SSBT corresponds to the choice of the second particle from the list of all particles $j = 1, \dots, j \neq i, \dots, N^{(l)}$ except $i$ , $k' = N^{(l)} - 1$.

The SGBT operator is like that of SSBT, except that the process is applied to the $N_{sel}$ particle in the corresponding cell rather than trying all particles in the cell. The stochastic interpretation of each of these operators gives a scheme of collision process in each cell $l$ with a collision probability of any randomly chosen particle pair $(i, j)$ as shown in Table 1:

**Table 1.** Probability function according to collision scheme.

| scheme | probability function |
|---|---|
| SBT | $w_{i,j} = \frac{(N^l - i) F_{num} g_{i,j} \sigma_{i,j} dt}{V^l}$ |
| GBT | $w_{i,j} = \frac{N^{(l)}(N^{(l)} - 1)}{N_{sel}(2N^{(l)} - N_{sel} - 1)} (N^{(l)} - i) \frac{F_{num} g_{i,j} \sigma_{i,j} dt}{V^{(l)}}$ |
| SSBT | $w_{i,j} = \frac{(N^l - 1) F_{num} g_{i,j} \sigma_{i,j} dt}{2V^l}$ |
| SGBT | $w_{i,j} = \frac{N^l (N^l - 1) F_{num} g_{i,j} \sigma_{i,j} dt}{N_{sel} \times 2V^l}$ |

## 3. Evaluation of the Collision Models

This section presents a detailed evaluation of various Bernoulli Trial (BT)-based collision partner selection algorithms, namely SBT, GBT, SSBT, and SGBT, as well as the NTC scheme, by applying them to several fundamental gas dynamic problems. These include the collision frequency benchmark, the homogeneous relaxation process characterized by the Bobylev–Krook–Wu (BKW) analytical solution [29–31], lid-driven microcavity flow [32–34], and hypersonic flow over a circular cylinder [35-36]. As part of the analysis, key metrics such as the collision frequency ratio (i.e., the ratio of numerical to analytical collision frequencies) and the fourth-order moment of the velocity distribution function (VDF) for the homogeneous relaxation case are examined and compared against theoretical predictions and the results obtained using the NTC scheme. All collision schemes have been implemented within an in-house DSMC Python code or modified versions of Bird's DSMC2 and DS2V codes [37-38] to facilitate consistent and accurate numerical simulations.

### 3.1 Collision frequency

The first test case involves calculating the equilibrium collision frequency ratio, which is the ratio of the numerical collision frequency to the theoretical one. The problem considered here is a spatially homogeneous monoatomic gas released from a random state to reach equilibrium. It should be noted that to demonstrate the differences between the algorithms, we extend the previous examination by obtaining results using a very small number of particles per cell (PPC) magnitudes. Here, we provide the results for recently proposed SGBT schemes and present a comprehensive comparison of all schemes. The test case is a spatially homogeneous monatomic gas with a reference particle diameter of 0.35 nm and molecular mass of $5\times10^{-26}$ Kg at a reference temperature of 300 K. The gas, with an initial number density of $n_{init}=1\times10^{20}$ $m^{-3}$, is released from a random state to reach equilibrium. The theoretical equilibrium collision rate per molecule ($CF_{th}$) is given by:

$$CF_{th} = 4nd^2\sqrt{\frac{\pi K_B T_{ref}}{m}}\left(\frac{T}{T_{ref}}\right)^{1-\omega}, \tag{17}$$

, where $n$ denotes the number density, $d$ is the gas molecular diameter, $K_B$ represents the Boltzmann constant, $T_{ref}$ is the reference temperature, $m$ represents molecular mass, and ω is the viscosity-temperature exponent.

The collision frequency in numerical predictions ($CF_{num}$) is calculated using the following equation:

$$CF_{num} = \frac{2N_{coll}}{N_P Time}, \tag{18}$$

, where $N_{coll}$ represents the total number of collisions occurring within each cell, $N_P$ is the average number of particles per cell. The $CF_{ratio}$ is derived by dividing a numerical value by its theoretical counterpart and at equilibrium, it must be equal to unity.

Figure 2 illustrates the results of the collision frequency benchmark test obtained using various BT-based collision partner selection schemes, in comparison with the conventional NTC method. The data were generated under varying conditions, including different particle counts per cell (PPC) and cell counts within the computational domain. For each PPC value, the time step was selected to ensure that the non-dimensional ratio of cell size to time step remained within the recommended range of 1-2, as suggested in Ref. [22]. All simulations in this section are carried out using a total of 1000 particles. The particles per cell (PPC) and the time step are adjusted such that the non-dimensional ratio of spatial step to time step, *dx/dt*, is consistently maintained at 1.

As shown in Figure 2, all BT-based schemes maintain the collision frequency ratio ($CF_{ratio}$) close to unity for all average particle numbers, which corresponds to the expected equilibrium value, even as the PPC decreases below one. In contrast, the NTC scheme exhibits a noticeable deviation from this behavior for all specified average particle values below one. This suggests that BT-based methods can deliver accurate results even with a reduced number of particles per cell, i.e., less than 1.

The standard NTC scheme fails to maintain the correct collision frequency at low PPC counts primarily due to its method of estimating the maximum collision product, $(\sigma_T g_r)_{max}$. In classic implementations, this value is often initially in a cell at a fixed, pre-calculated constant based on bulk gas properties, such as the initial temperature. While this approach works in high PPC conditions, it becomes unreliable in low-PPC scenarios where the statistical representation is poor. In a cell with very few particles, a randomly selected pair can have an actual relative velocity that is significantly higher than the pre-estimated maximum. When this happens, the algorithm systematically underestimates the number of candidate pairs that should be checked for collision (see Eq. 19), causing the simulated numerical collision frequency to drop below the correct theoretical value, as observed in Fig. ٢. In the next section, we introduce a NTC-Pre-Scan method that resolves this issue by dynamically calculating a localized $(\sigma_T g_r)_{max}$ for each cell at every timestep through a quick sampling, ensuring the value is always representative of the actual particles present and thereby maintaining the correct collision frequency regardless of the PPC.

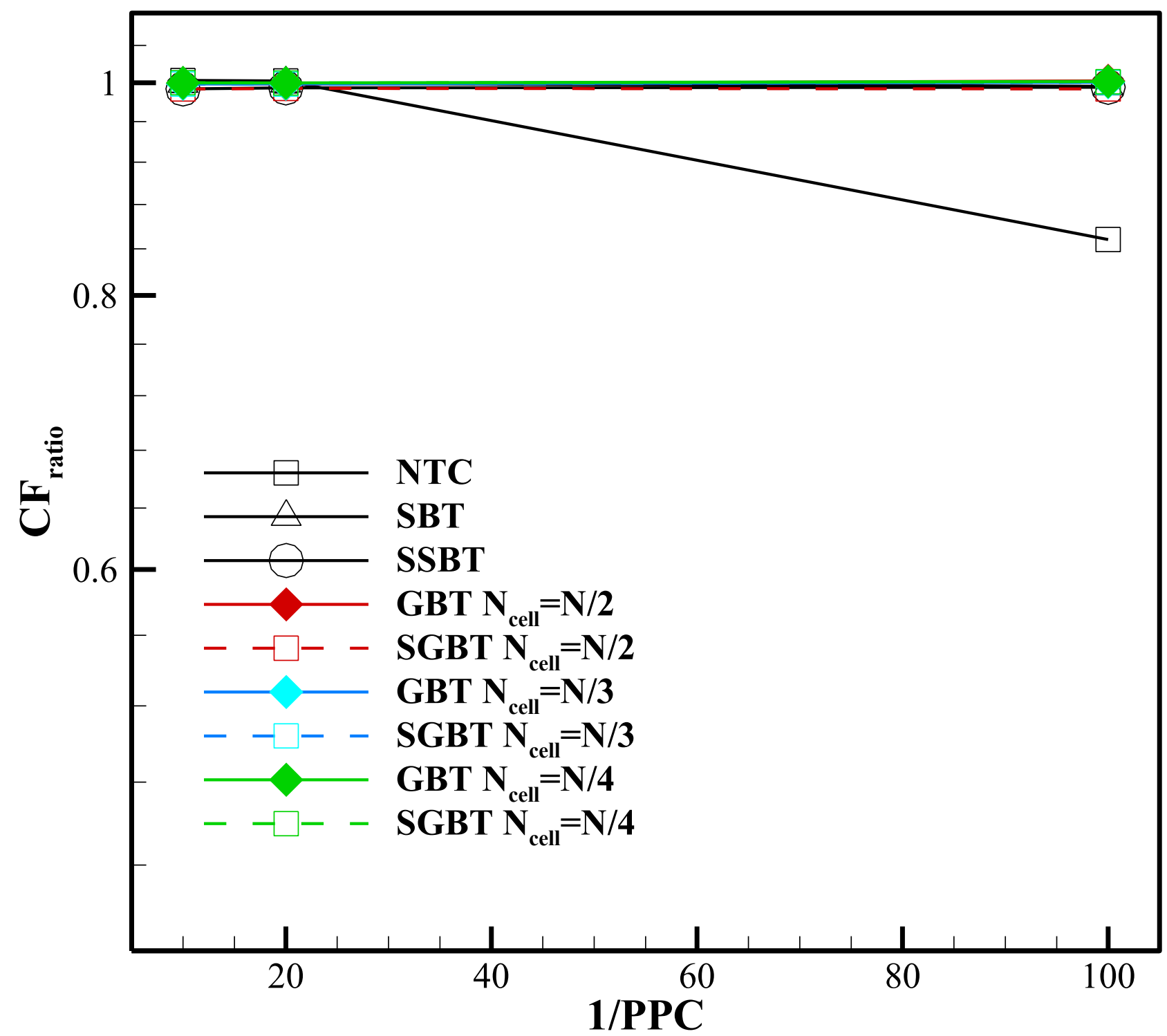

Fig 2. Comparison of $CF_{ratio}$ with different collision partner selection schemes.

### 3.1.1 Advancement in the NTC scheme to improve collision frequency

### A) Numerical DSMC Solver in Python

As the discussions in the previous section demonstrate, the standard NTC collision scheme is unable to capture the correct collision frequency at low PPC. To resolve this issue, we propose a new variant of the NTC scheme that utilizes a Markovian "Pre-Scan" algorithm, along with some other minor improvements, as detailed below. To implement the new algorithm, we developed a Python script that is a sophisticated physics simulation tool designed to model a 0-D homogeneous gas relaxation problem using the DSMC method. The primary purpose is not to simulate a complex geometry, but to serve as a numerical laboratory for comparing collision algorithms. It implements several different statistical collision schemes to test their accuracy and physical fidelity. Additionally, we utilize the code for analyzing statistical physics, which performs advanced statistical analysis on the simulated gas to ensure it accurately reproduces the behavior of an ideal gas at equilibrium, as described by statistical mechanics. The code features a robust system for handling simulations with very few particles per cell (i.e., "fractional particles"), a challenging area in DSMC research. The simulation starts with particles having a non-equilibrium velocity distribution and tracks their evolution until they "relax" into the theoretical Maxwell-Boltzmann distribution for the given temperature.

As the code utilizes various collision models, including BT-family members, we implemented an intelligent parameter calculation and mesh refinement strategy. This is the code's most innovative feature, designed to handle the "probability exceed" problem. In DSMC, the probability of any single pair colliding should be less than 1. In simulations with very few particles per cell, it becomes difficult to choose a cell size (*dx*) and time step (*dt*) that satisfy this condition without violating other physical constraints. Our algorithm automatically refines the simulation mesh to ensure the probability of collision remains below 1, thereby guaranteeing a physically valid simulation. The "Iterative Refinement Algorithm" is as follows:

1. Heuristic Link: The code enforces a key relationship: $dx/\lambda \approx dt/\tau_c$ (or $dx/dt \approx 1$ in dimensionless terms), where $\lambda$ is the mean free path and $\tau_c$ is the mean collision time. This links the spatial resolution (*dx*) directly to the temporal resolution (*dt*). This condition is equivalent to having CFL=1 in conventional CFD, as shown in the previous works of Roohi and co-workers, i.e., [22], which should be satisfied for the BT collision schemes. Note that this constraint is in addition to the spatial constraint of $dx/\lambda < 1/3$ and the temporal constraint $dt/\tau_c < 1/3$.

2. Initial State: The simulation starts with a base number of cells.

3. Check Probability: It calculates the maximum expected collision probability ($P_{max}$) based on the current number of cells and the corresponding time step. Note that the code does not know the actual speed of the molecules at startup time. So, the value of $P_{max}$ should be set by trial and error.

4. Decision:

- If $P_{max}$ is below the target, the parameters are considered good, and the process stops.
- If $P_{max}$ is too high, the algorithm aggressively increases the number of cells.

5. Loop and Converge: Because of the heuristic link in step 1, an increasing number of cells makes *dx* smaller, which in turn forces *dt* to become smaller. A smaller *dx* (cell size) and a smaller *dt* both work to drastically reduce the collision probability in the next iteration. This loop continues until $P_{max}$ falls below the target threshold.

This automated "fine mesh refinement" allows the code to run accurately even in challenging fractional-particle regimes, trading higher computational cost for physical correctness. The code, in addition to evaluating the collision frequency ratio, checks if the random variation in the number of particles per cell follows the theoretical Poisson distribution. Temporal and spatial correlation functions check that the simulation does not create artificial "memories" or spatial relationships between particles. In a properly randomized gas at equilibrium, these correlations should quickly decay to zero, which the generated plots verify.

**B) Markovian NTC Algorithm**

In the traditional implementation of the NTC scheme, the number of selections is proportional to $0.5 \times N \times \langle N \rangle$, where $N$ is the instantaneous number of particles and $\langle N \rangle$ is the time-averaged number of particles in that cell. In the modern implementation, following Bird [38], our Python code uses the more common contemporary formula, which is proportional to $0.5 \times N \times (N - 1)$.

The second improvement is on the handling of fractional pairs in the NTC scheme. Even modern available DSMC solvers, which implement the NTC schemes, use a deterministic remainder for $A_{\text{sel}}$ formula remainder. The total number of candidate pairs to be selected for collision checks within a single cell over the time step $\Delta t$ is calculated as:

$$A_{\text{sel}} = \frac{N_c(N_c - 1)}{2} \frac{F_{\text{num}}(\sigma_T g_r)_{\max} \Delta t}{V_c} \tag{19}$$

, where $N_c$ is the number of simulator particles currently in the cell, $F_{\text{num}}$ is the number of real gas particles represented by each simulator particle, $V_c$ is the volume of the cell, $\Delta t$ is the simulation time step, $(\sigma_T g_r)_{\max}$ is the majorant product, representing the maximum value of the product of the total collision cross-section ($\sigma_T$) and the relative speed ($g_r$) between any two particles in the cell.

Eq. (19) gives a real number, while the number of checked pairs for collisions should be an integer. In traditional and modern Bird code and solvers, i.e., [39-40], the fractional part of the number of selections is stored in a variable and explicitly added to the calculation in the next time step. Our Python code utilizes a probabilistic method known as “Probabilistic Rounding”. The fractional part is treated as a probability of selecting one additional pair. This is achieved by adding a random fraction ($R_f$) before converting the number to an integer: int($A_{\text{sel}}$+ $R_f$). This could be a new technique in modern DSMC codes using the NTC scheme.

Once a candidate pair of particles is randomly selected in the NTC scheme, the probability that this pair undergoes a collision is given by the ratio of its actual collision product to the majorant product:

$$P_{\text{coll}} = \frac{(\sigma_T g_r)_{\text{actual}}}{(\sigma_T g_r)_{\text{max}}} \tag{20}$$

The third improvement in our algorithm is on the maximum collision product, $(\sigma_T g_r)_{\text{max}}$ , which proved crucial in improving the $CF_{Ratio}$ prediction of the NTC scheme, especially when the number of particles per cell is small. Both old and modern Bird's DSMC code initialize the maximum collision product, $(\sigma_T g_r)_{\text{max}}$, using a hardcoded, fixed estimate rather than calculating it dynamically from the initial particle velocities. In the Bird implementation, this value is determined by a formula that combines a reference collision cross-section with a guessed velocity and a scaling factor for the initial temperature, or just a guessed velocity, such as the most probable speed on the cells. This approach serves as a pragmatic, albeit simplistic, starting point intended to get the simulation up and running. This approach yields an NTC simulation of the order of $N_c$, but our investigation has shown that this is the source of the error in the NTC scheme when $N_c$ is small.

The suggested "Pre-Scan" algorithm is a preliminary optimization step employed in majorant frequency-based collision schemes such as the NTC, Majorant Frequency Scheme (MFS), and Nearest Neighbor (NN) methods. Before selecting the final collision pairs within a cell for a given timestep, the algorithm first randomly samples several particle pairs. For each sampled pair, it computes the product of the collision cross-section and the relative velocity, denoted as $\sigma_T g_r$. The primary objective of this process is to find a reliable, localized estimate for the maximum possible value of this parameter, $(\sigma_T g_r)_{\max}$, within that specific cell at that particular time step. The number of samples taken during this Pre-Scan is adaptively adjusted based on the particle population of the cell, which prevents unnecessary computational work in sparse cells.

The primary advantage of this technique is a significant increase in simulation accuracy, particularly when PPC is small. Methods like NTC rely on the $(\sigma_T g_r)_{\max}$ value to determine the total number of candidate collision pairs to test, Eq. (19). In addition, if this value is set too high, a large fraction of candidate collisions will be rejected, see Eq. (20), leading to wasted computational effort. Conversely, if the value is set too low, the simulation becomes physically inaccurate because it will fail to consider all valid collisions, i.e, small $A_{sel}$. By performing a quick "Pre-Scan" to obtain a local and current estimate of $(\sigma_T g_r)_{\max}$, the algorithm ensures that the subsequent collision selection process is both physically correct and computationally efficient, eliminating the need to use a single, global, and overly conservative maximum value for the entire simulation domain.

A stochastic process is defined as Markovian if it possesses the "memoryless" property. This central tenet of probability theory dictates that the probability of transitioning to a future state depends solely on the current state of the system, not on the sequence of events or states that preceded it. In the context of the DSMC collision step within a single computational cell, this principle has a precise physical interpretation. The probability of a collision occurring during a time step should be a function only of the current ensemble of particles—their number, positions, and velocities—within that cell at that specific moment. The process should not depend on the state of the cell at previous time steps or on global parameters set at the beginning of the simulation. Adherence to this property is critical for correctly modeling the principle of molecular chaos, where each collision event is treated as statistically independent of the system's history. The standard NTC method, as implemented in many widely used codes, violates the Markov property. This violation stems from its reliance on the majorant product. As was described earlier, this value is often initialized using a "hardcoded, fixed estimate" based on bulk gas properties at the simulation's outset. This initial value introduces a form of system "memory" that persists throughout the simulation, leading to a non-Markovian process.

The chain is as follows:

The number of candidate collision pairs to be checked, $A_{sel}$, is calculated using Equation (19), which is directly proportional to this fixed $(\sigma_T g_r)_{\max}$. This $(\sigma_T g_r)_{\max}$ value, determined from initial conditions, is carried forward, acting as a memory of a past state. In simulations with strong gradients, such as hypersonic shock waves, the local gas properties within a cell can undergo significant changes. The temperature can increase significantly, causing the actual maximum $\sigma_T g_r$ among the particles currently in the cell to exceed the pre-set, "remembered" $(\sigma_T g_r)_{\max}$. Consequently, the calculation of $A_{sel}$ becomes dependent on a past state (the initial condition) rather than the present state (the actual particles in the cell). This is a direct violation of the memoryless principle that defines a Markovian process.

The "Probability Exceed Ratio" metric introduced later in the manuscript is not merely a measure of numerical stability; it serves as a direct diagnostic for this non-Markovian flaw. When the actual $\sigma_T g_r$ of a selected pair is greater than the stored $(\sigma_T g_r)_{\max}$, it signifies that the system's memory of the "maximum" is outdated and incorrect. The standard NTCs' struggle with a high probability exceed ratio, as empirically demonstrated later in Figures 4b and 5b of the current manuscript, is evidence of the system attempting to operate outside the bounds set by its past state—a hallmark of this memory-induced error. This frames the failure of the standard NTC not just as a numerical inaccuracy but as a fundamental breakdown of its underlying stochastic model.

The proposed NTC-Pre-Scan algorithm directly rectifies this theoretical flaw. As explained in this Section of the manuscript, the key mechanism is that at the beginning of every collision step in every cell, the algorithm performs a quick, adaptive sampling of the currently present particles to determine a localized, time-accurate $(\sigma_T g_r)_{\max}$. By discarding the old, global $(\sigma_T g_r)_{\max}$ and recalculating it from the present state, the algorithm effectively "erases" the system's memory of past conditions. The subsequent calculation of $A_{sel}$ (Eq. 19) and the collision probability $P_{\text{coll}}$ (Eq. 20) is now conditioned exclusively on the current particle ensemble in the cell. The future state (the collision outcome) is rendered independent of the past, thereby satisfying the Markov property.

The NTC-Pre-Scan algorithm demonstrates superior performance over the standard NTC method without deteriorating the accuracy at low PPC. Using our developed Python code, the simulation is initialized for Argon gas. As previously mentioned, the crucial numerical parameters are not fixed but are dynamically calculated to ensure physical accuracy based on user input. After the user specifies the desired average PPC, a setup function recalculates the total number of cells and the time step. This process often employs a "fine mesh refinement" strategy, significantly increasing the cell count for low-PPC scenarios to control numerical errors, while simultaneously adjusting the time step to maintain stability.

As shown in Figs. 3 where PPC is 20, the Pre-Scan algorithm is optimal, immediately achieving a perfect collision frequency ratio ($CF_{ratio}$ = 1.0) and a probability exceed ratio of zero *from the start of the simulations*. This contrasts with the standard NTC method, whose behavior depends on the initial maximum collision parameter, $(\sigma_T g_r)_{max}$, and still struggles with initial inaccuracies. As shown in the figure, various initial guesses for this value were considered, each of them influencing the start-up behavior of the frequency ratio. The Pre-Scan method's ability to find the correct local parameters from the start highlights its inherent accuracy and efficiency, i.e., the probability exceed ratio is almost zero from the start of the simulations. Note that this initial error can significantly impact unsteady problems with short simulation times.

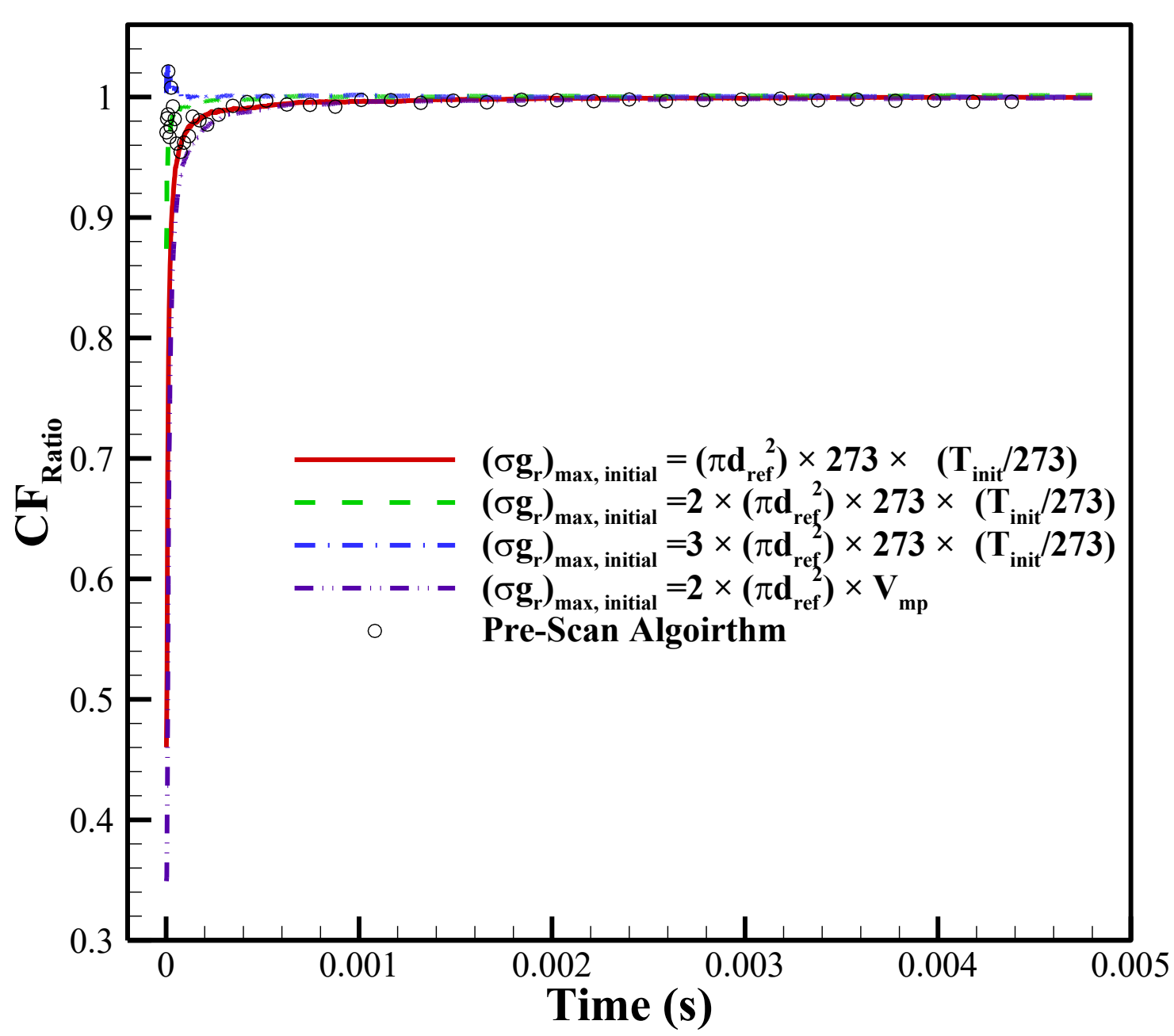

a) $CF_{Ratio}$

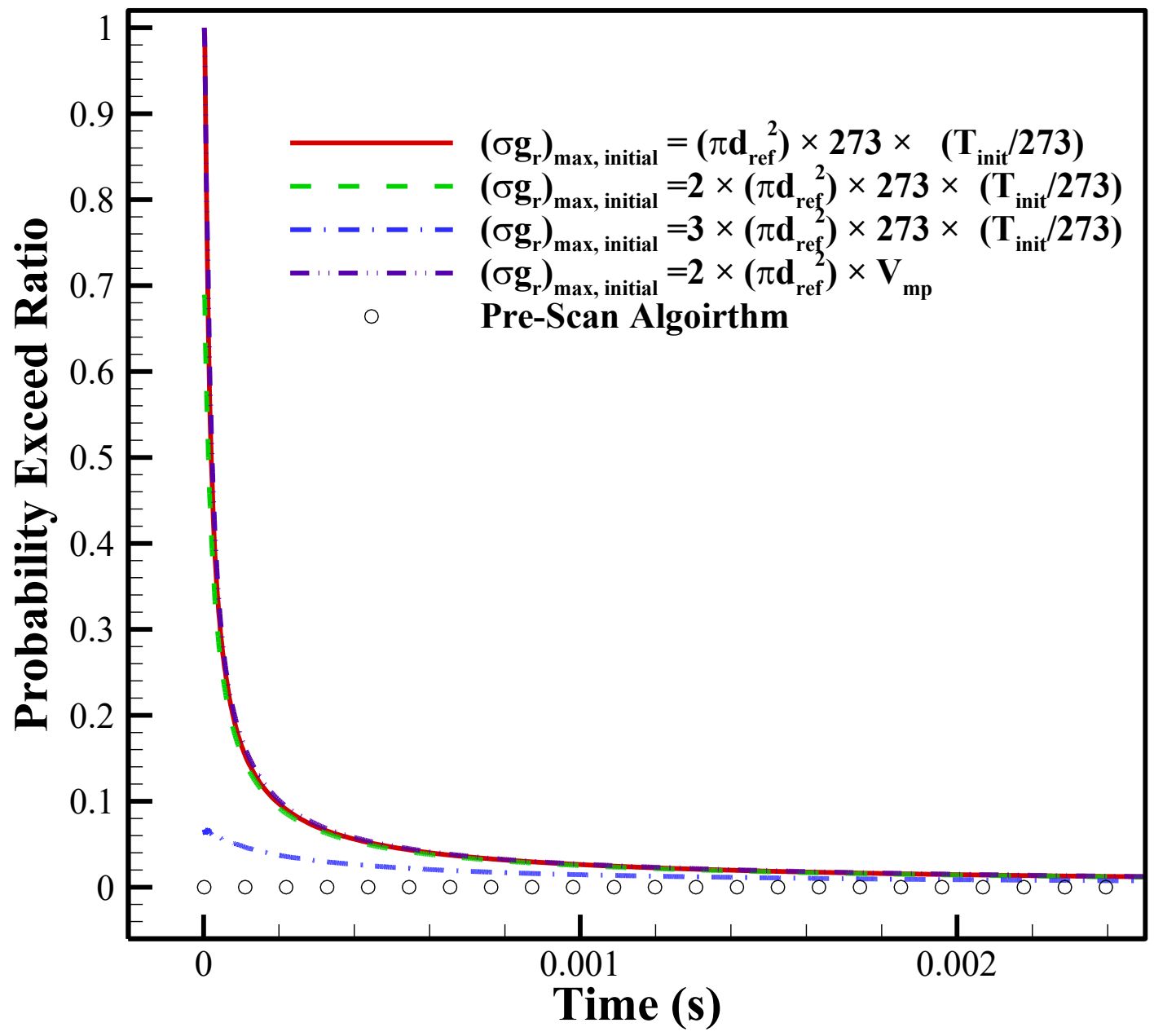

b) Probability Exceed Ratio

Fig 3. Comparison of $CF_{ratio}$ and exceed ratio in the NTC scheme, PPC=20.

The critical advantage of the NTC-Pre-Scan is most evident in challenging low-particle simulations, where the standard NTC method begins to fail (PPC = 2, as reported in Fig. 4, and PPC = 0.01, as considered in Fig. 5). Note that PPC=0.01 means that we have an average of 0.01 particles per cell; i.e., if we have 100 cells, we have a total of 10 particles in the domain. These particles move within the domain, and when two particles are in the same cell, the collision possibility is checked. Considering this, it can be concluded that capturing the correct frequency ratio at low PPC is quite challenging and needs a very accurate collision algorithm.

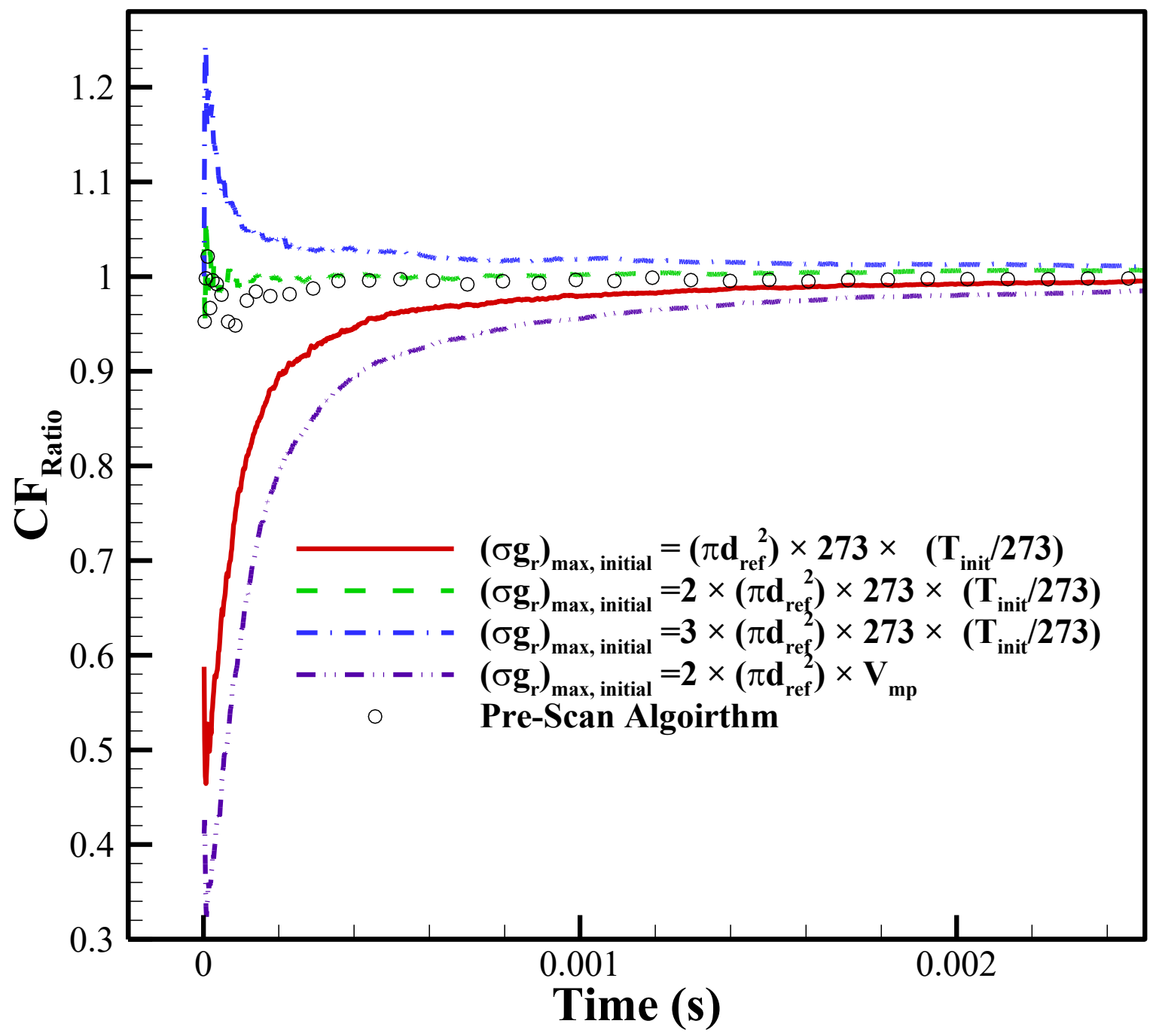

a)

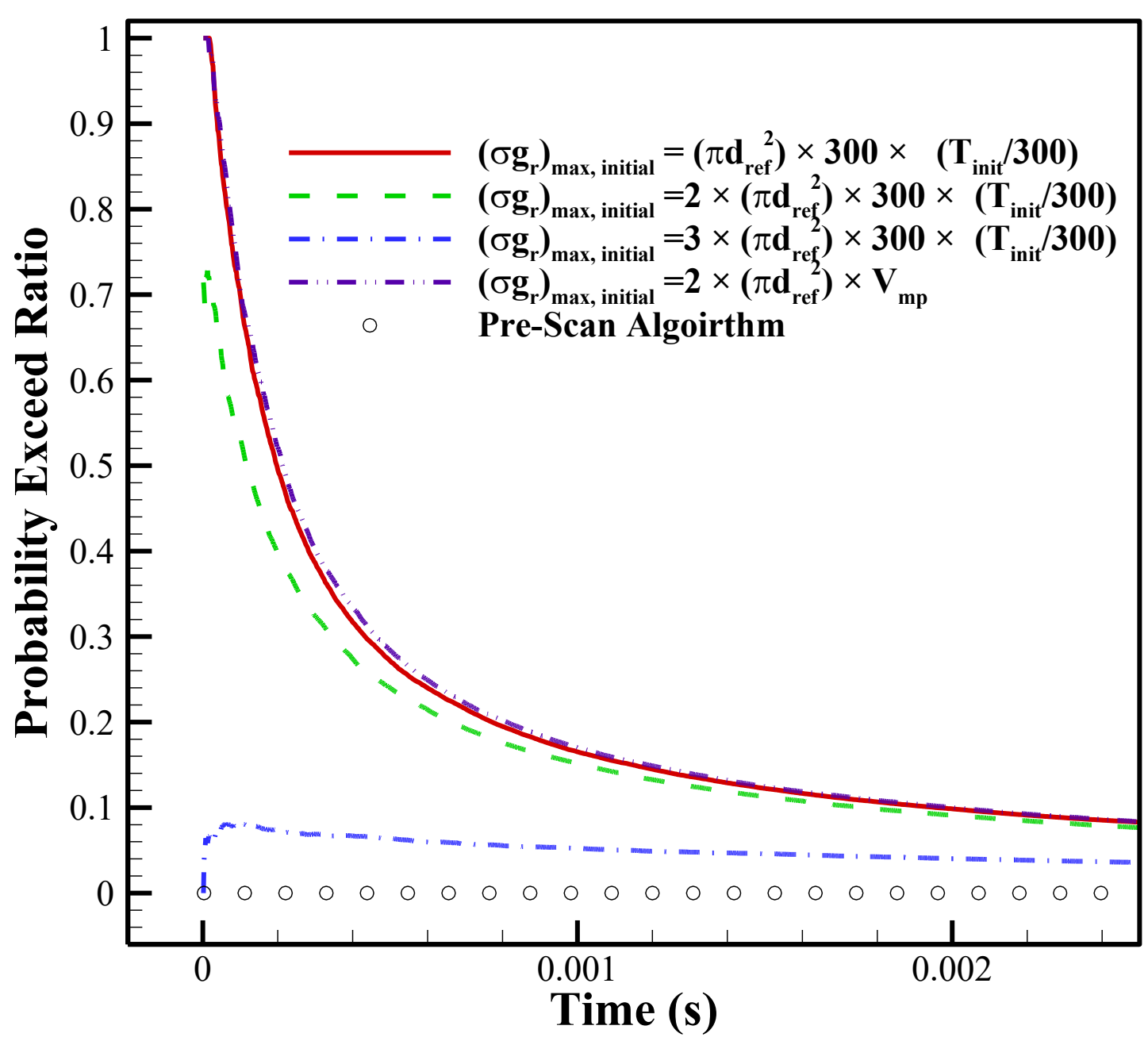

b)

Fig 4. Comparison of $CF_{ratio}$ and exceed ratio in the NTC scheme, PPC=2.

The standard method yields a physically incorrect result at PPC=0.01, capturing only half the necessary collisions ($CF_{ratio} \approx 0.5$) and suffering from a 100% probability exceed. In stark contrast, the NTC-Pre-Scan performs remarkably, maintaining a perfect CF ratio of 1.0 and zero exceedances. This success stems from its adaptive ability to determine the correct local collision dynamics from the few particles present, proving it is a far more robust and reliable algorithm for accurate DSMC simulations across any gas density. Note that, although the computational costs of the Pre-Scan algorithm are higher than those of the standard NTC algorithm, we recommend using it in low PPC conditions, where the number of additional operations is negligible, as there are few particles per cell.

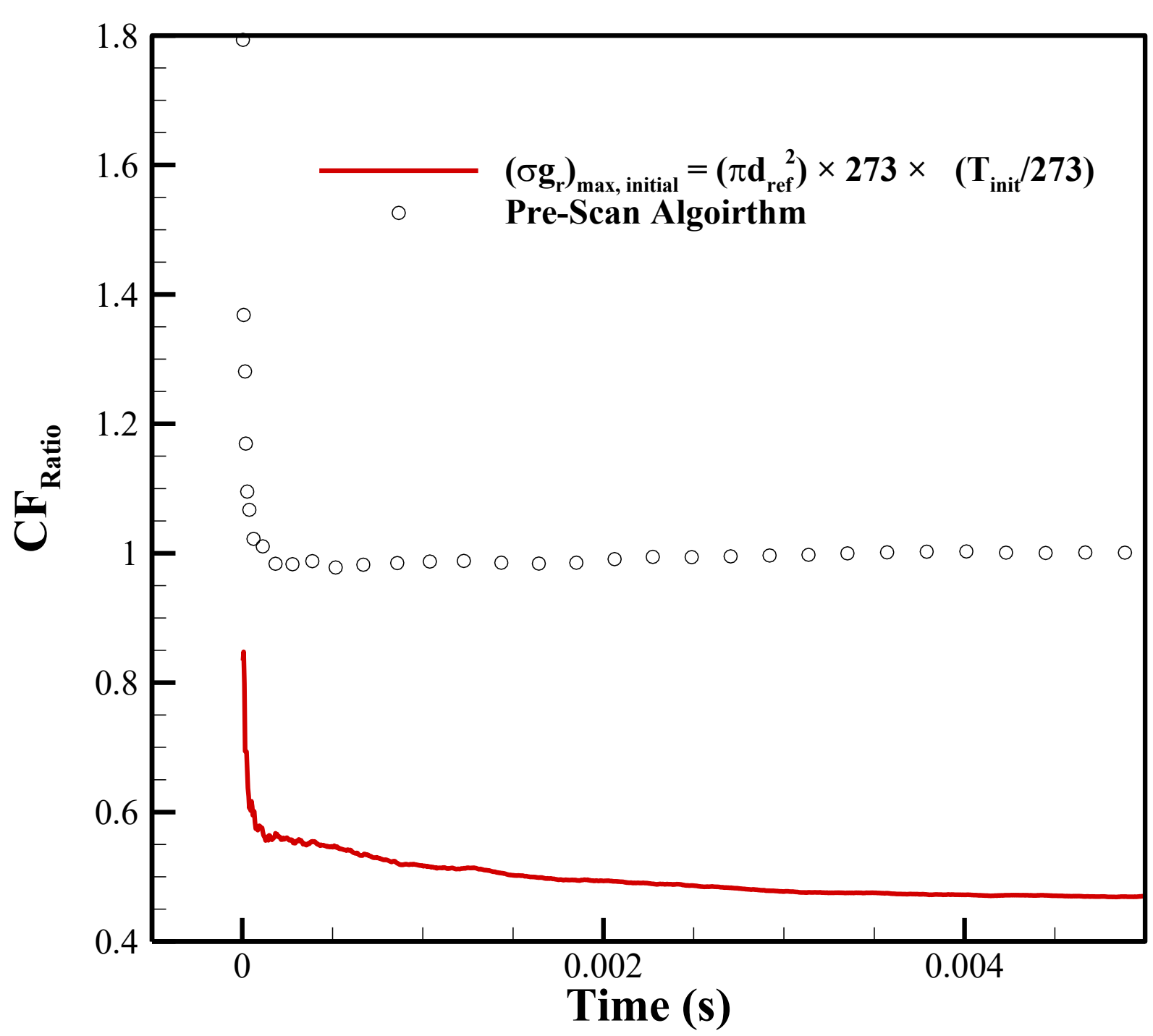

a) $CF_{Ratio}$

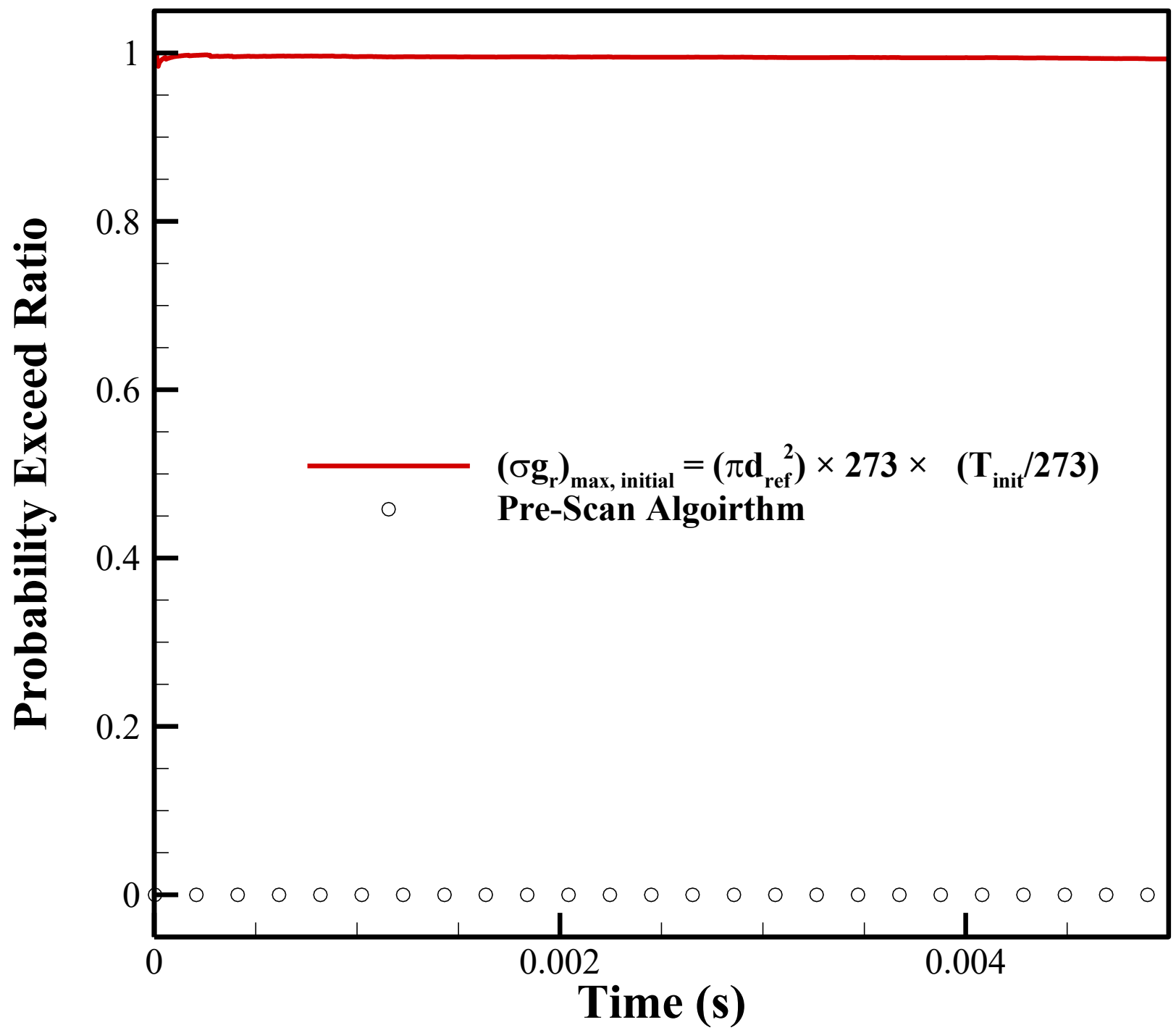

b) Probability Exceed ratio

Fig 5. Comparison of $CF_{ratio}$ and exceed ratio in the NTC scheme, PPC=0.01.

Fig. 6 presents a comparative analysis of the performance and accuracy of the NTC-Pre-Scan and SBT collision schemes. The results were obtained using our Python code with PPC=0.01. Both methods, despite using very low PPC, demonstrate success in achieving thermal equilibrium, as evidenced by their final speed distributions, which show excellent agreement with the theoretical Maxwell-Boltzmann distribution at the final temperature of 281.3 K. Furthermore, the collision frequency ratio for both schemes fluctuates tightly around the ideal value of 1.0, indicating that both methods correctly compute the macroscopic collision rate over time. Both schemes show a collision separation (SOF) as small as 0.01, which is quite acceptable for a high-fidelity simulation.

A distinction arises in the probability exceed ratio, which measures the percentage of selected collision pairs with a calculated collision probability greater than one. The NTC-Pre-scan scheme maintains a probability exceed ratio of zero; the SBT scheme (b) exhibits random spikes where the collision probability exceeds 1, reaching as high as 5%. This phenomenon is a numerical artifact that can compromise the physical fidelity of the BT family schemes and should be avoided and kept as low as possible. This can be achieved by using a smaller time step in the SBT simulations, i.e., by adjusting the expected collision probability ($P_{max}$) in our Python solver to ensure no exceedance in the probability.

Fig. 7 presents an advanced statistical analysis for the (a) NTC-Pre-Scan and (b) SBT collision schemes, both operating under the challenging condition of a very low PPC = 0.01. The results demonstrate that both methods accurately capture the fundamental physics of a gas in thermal equilibrium, even in this sparse-data environment. As previously discussed and illustrated in Fig. 2, the standard NTC cannot exhibit the behavior depicted in Fig. 7a.

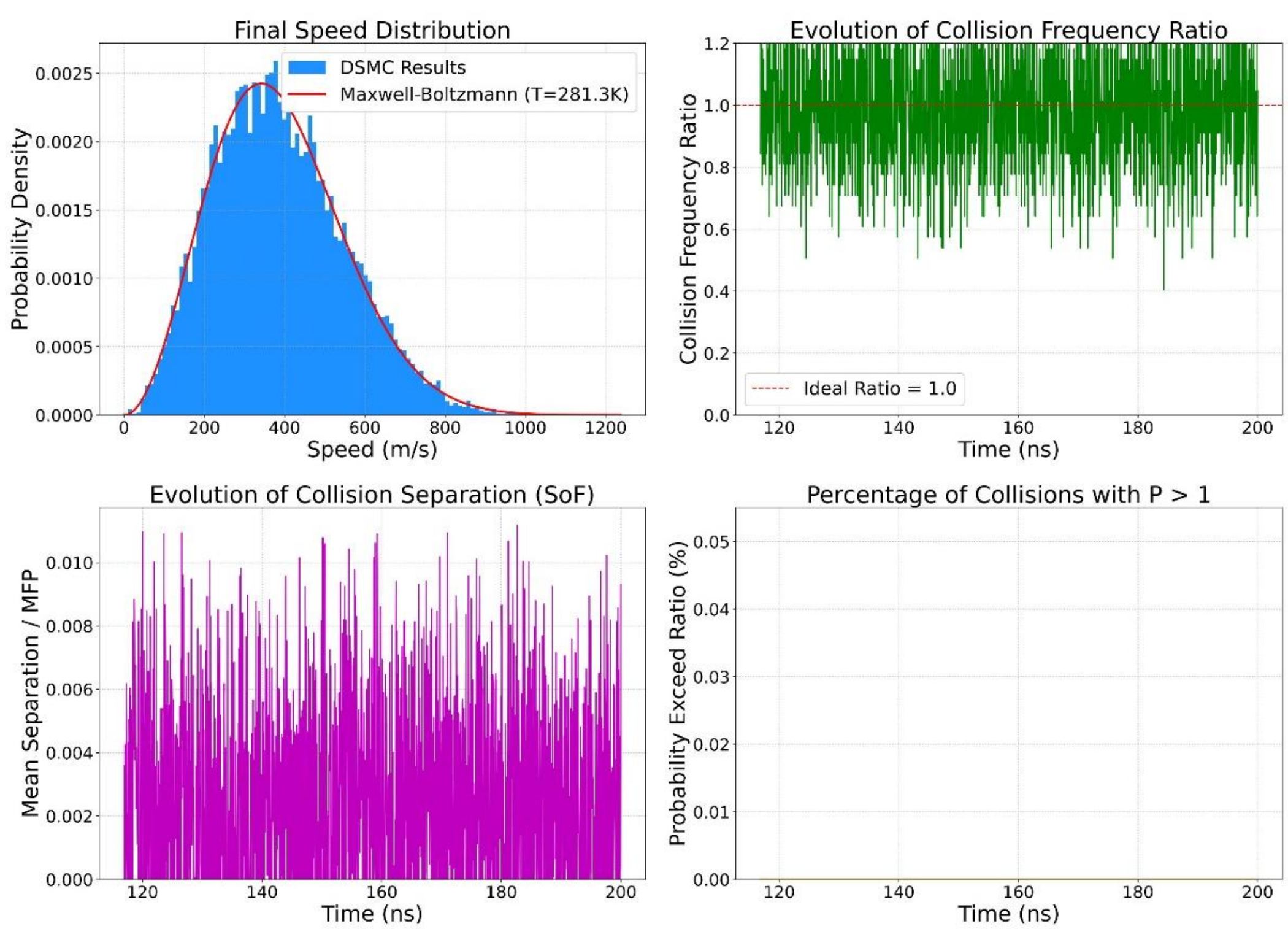

a) NTC

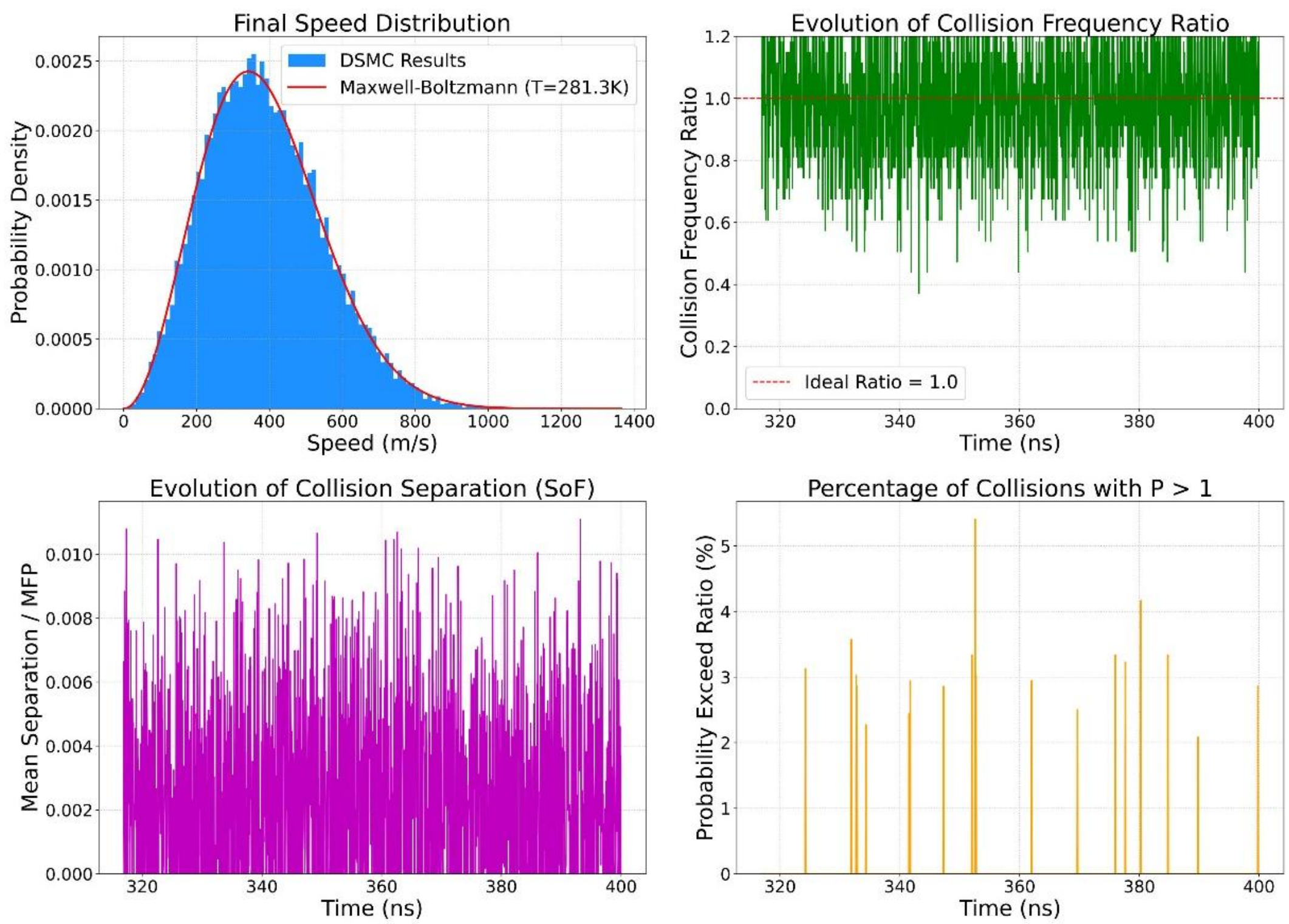

b) SBT

Fig. 6: Speed distribution, frequency ratio, SOF, and probability exceed for (a) NTC scheme and (b) the SBT scheme.

Number density fluctuation is shown in the top plots in Fig. 7-a-b. This plot shows the probability distribution of the number of particles found in a simulation cell. For both NTC and SBT, the simulation results (blue histogram) show a perfect match with the theoretical Poisson distribution (red dashed line), as confirmed by a $\chi^2$ value of 0.000. This indicates that both algorithms correctly model the random spatial distribution of particles, a cornerstone of statistical mechanics. Temporal auto-correlation, shown in the Middle of Fig. 7-a-b, measures how gas properties (velocity, temperature) in a cell are correlated with themselves over time. In both figures, the correlation for all properties drops to zero almost instantly for any non-zero time lag. This signifies the absence of unphysical "memory" effects and confirms that the simulations correctly adhere to the principle of molecular chaos, where the state of the system at one moment is independent of its state at the next. Spatial Correlation, shown in the bottom row of Fig. 7-a-b shows the correlation of properties between a central cell and its neighbors. For both methods, a sharp peak exists only at zero separation (a cell correlated with itself), with zero correlation between different cells. This confirms that no unphysical spatial structures or dependencies exist, further validating the assumption of molecular chaos.

Note that we performed similar calculations for GBT, SSBT, and SGBT collision schemes with low PPC, and all showed the same behavior as shown in Fig. 7-b for the SBT scheme. The only parameter to consider in BT family schemes is to keep the probability exceed ratio as low as possible; otherwise, the collision frequency ratio and other statistical properties of the simulations will deteriorate. Additionally, we should note that the implementation of the Pre-scan algorithm does not resolve the issue of repeated collisions in the NTC at low PPC, which was discussed in detail by Stefanov in [17]. In a zero-dimensional (0D) relaxation-to-equilibrium problem, repeated collisions between the same pair of particles are considered a benign statistical artifact that does not impact the overall collision frequency. This is because the system is perfectly homogeneous, meaning all simulated particles exist within a single conceptual cell without any spatial gradients or particle movement steps. The selection of colliding pairs is a purely random sampling process based on the statistical properties of the entire particle ensemble. After a collision, the particles' new velocities are returned to the global pool, and the system effectively has no memory of that specific interaction. Consequently, the random re-selection of the same pair in a subsequent step does not bias the macroscopic collision frequency, which is an averaged property calculated from all possible pairings in the system. In stark contrast, repeated collisions become a significant source of numerical error in multi-dimensional simulations (1D, 2D, or 3D) that involve spatial gradients. In these cases, the principle of locality is paramount, dictating that only particles that are physically proximal should collide. A repeated collision, particularly between particles that are spatially distant within the same cell, violates this principle, leading to a non-physical transport of momentum and energy across the cell and introducing inaccuracies in the calculation of critical transport properties, such as viscosity and thermal conductivity. Furthermore, it creates an artificial statistical correlation between the colliding pair, which undermines the fundamental assumption of molecular chaos upon which the DSMC method is built. Therefore, while irrelevant to the bulk collision rate in a 0D relaxation problem, minimizing repeated collisions is crucial for achieving physically accurate results in simulations with spatial variations. This is the benefit of the BT schemes, as they naturally control repeated collisions. While the NTC-Pre-Scan algorithm successfully corrects the collision frequency and restores theoretical consistency, it does not address the separate issue of repeated collisions. As demonstrated by Stefanov [17], such collisions can introduce significant numerical errors in multi-dimensional simulations with spatial gradients. This limitation motivates our systematic evaluation of the Bernoulli-Trial family of schemes, which are specifically designed to eliminate this artifact.

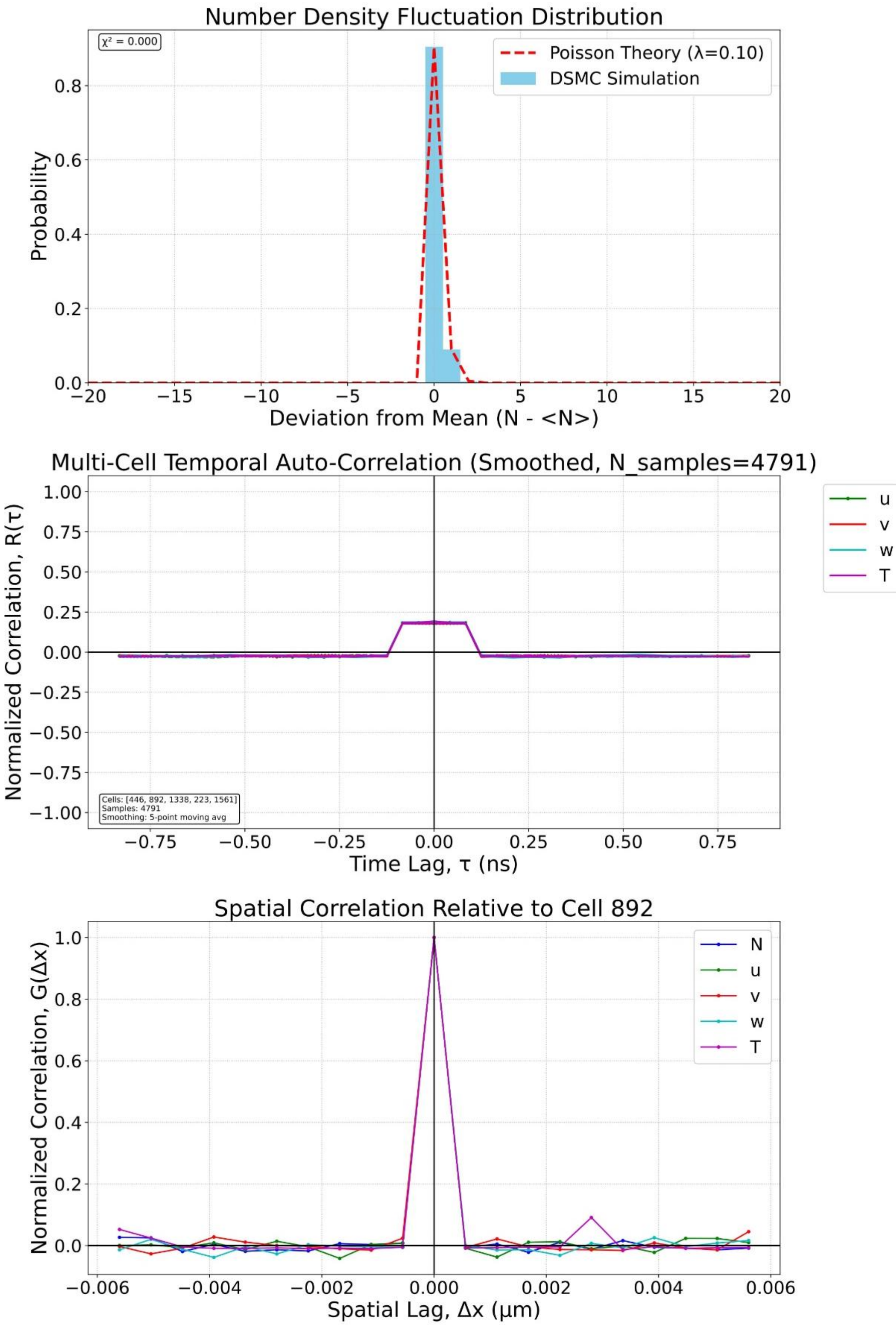

Number Density Fluctuation Distribution
χ² = 0.000
Poisson Theory (λ=0.10)
DSMC Simulation
Probability
Deviation from Mean (N - <N>)
Multi-Cell Temporal Auto-Correlation (Smoothed, N_samples=4791)
u
v
w
T
Normalized Correlation, R(τ)
Cells: [446, 892, 1338, 223, 1561]
Samples: 4791
Smoothing: 5-point moving avg
Time Lag, τ (ns)
Spatial Correlation Relative to Cell 892
N
u
v
w
T
Normalized Correlation, G(Δx)
Spatial Lag, Δx (μm)

a) NTC Scheme

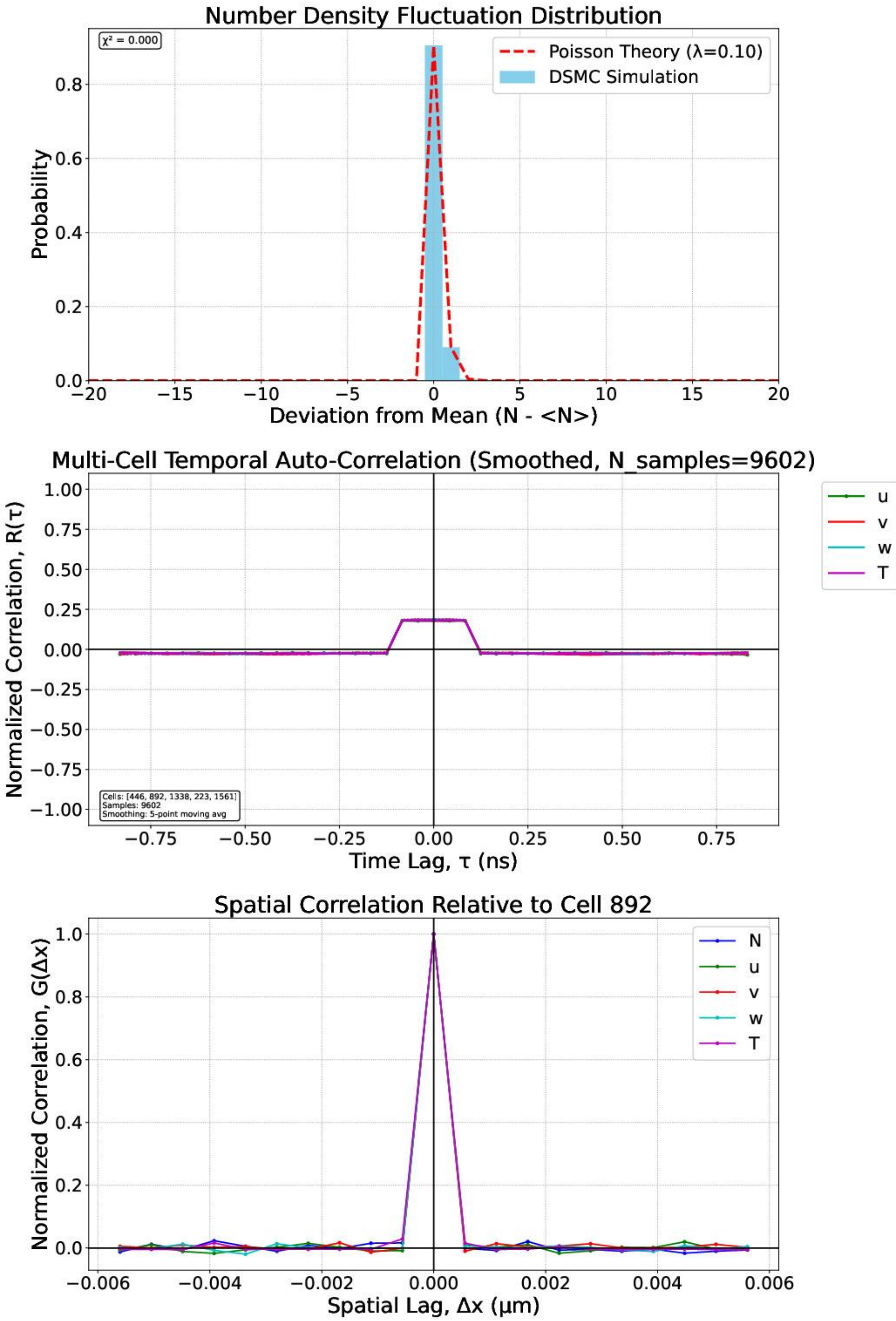

b) SBT Scheme

Figure 7: Report of number density fluctuation, temporal and spatial auto-correlations for NTC and SBT schemes, PPC=0.01

### 3.2 BKW Solution to the Boltzmann Equation

The homogeneous relaxation process to the unsteady Boltzmann equation, using the BKW solution, was studied by Shoja-Sani et al. with the NTC, SBT, and GBT schemes [41]. Here, the same problem is considered to evaluate the SSBT and SGBT schemes. To quantify the difference between the DSMC results and theory, the fourth moment of the x-velocity distribution function is used as the parameter and obtained as:

$$< c_x^4 >= 3\left(\frac{kT}{m}\right)^2 \frac{(1+2\beta)}{(1+\beta)^2} \tag{21}$$

, where $\beta$ is the nonequilibrium parameter ($0 \leq \beta \leq 2/3$), which vanishes at the equilibrium. Also, the average of the fourth moment of the three velocity components is considered. This reads as follow:

$$< c^4 >= \frac{< c_x^4 > + < c_y^4 > + < c_z^4 >}{3} \tag{22}$$

This study considers a Maxwell molecule characterized by a reference diameter of $4.17\times10^{-10}$ m at a reference temperature of 273 K and a molecular mass of $6.64\times10^{-26}$ kg. The gas is maintained at a temperature of 273 K with a number density of $1\times10^{20}$ $m^{-3}$. Employing the variable soft sphere (VSS) model with a scattering parameter $\alpha$ equal to $\sqrt{2}$ yields a Bobylev parameter of $\lambda_B = 5931.98$ $s^{-1}$. An initial $\beta$ value of 0.65 was chosen for the analysis.

It is worth noting that all the simulations presented here indicate the evolution of the fourth moment over 500 μs. The total number of particles is constant and equal to 400,000. The only exception is the PPC of 0.1, which we performed using 100000 total number of simulators. The number of cells in each set of simulations was adjusted to achieve the desired PPC magnitude. The study begins with a relatively large PPC magnitude, which gradually decreases, to examine the impact of mesh refinement on the performance of each collision model. The time step is selected to be smaller than the mean collision time of 38.5 μs. Additionally, the constraint imposed by the Bernoulli trials scheme is considered, ensuring that the proportion of collision probabilities exceeding a value of one remains below 1%.

Figure 8 presents the fourth moment of the velocity distribution function (VDF) for the homogeneous relaxation solution using the SSBT collision scheme. Specifically, Figure 8a illustrates the fourth moment of the x-velocity component of the VDF, computed using various PPC magnitudes, including PPC = 0.1, 2, 5, 10, and 40, compared with the theoretical prediction. As shown, even with decreasing PPC magnitudes, the SSBT scheme consistently maintains a high level of agreement with the theoretical solution throughout the entire simulation period. The zoomed view in this figure shows more details of the SSBT schemes' behavior in simulation with different PPC magnitudes. Figure 8-b demonstrates good agreement between the theoretical predictions and the simulation results for the average fourth moment across all three velocity components. This consistency confirms that the SSBT scheme is capable of accurately capturing higher-order moments of the velocity distribution function, even when utilizing a relatively small number of particles per cell. Excellent agreement in the fourth moment of the x-velocity component ($c_x^4$) and average fourth moment across all three velocity components ($c^4$) is obtained with the SSBT scheme, like previous studies on collision schemes [41].

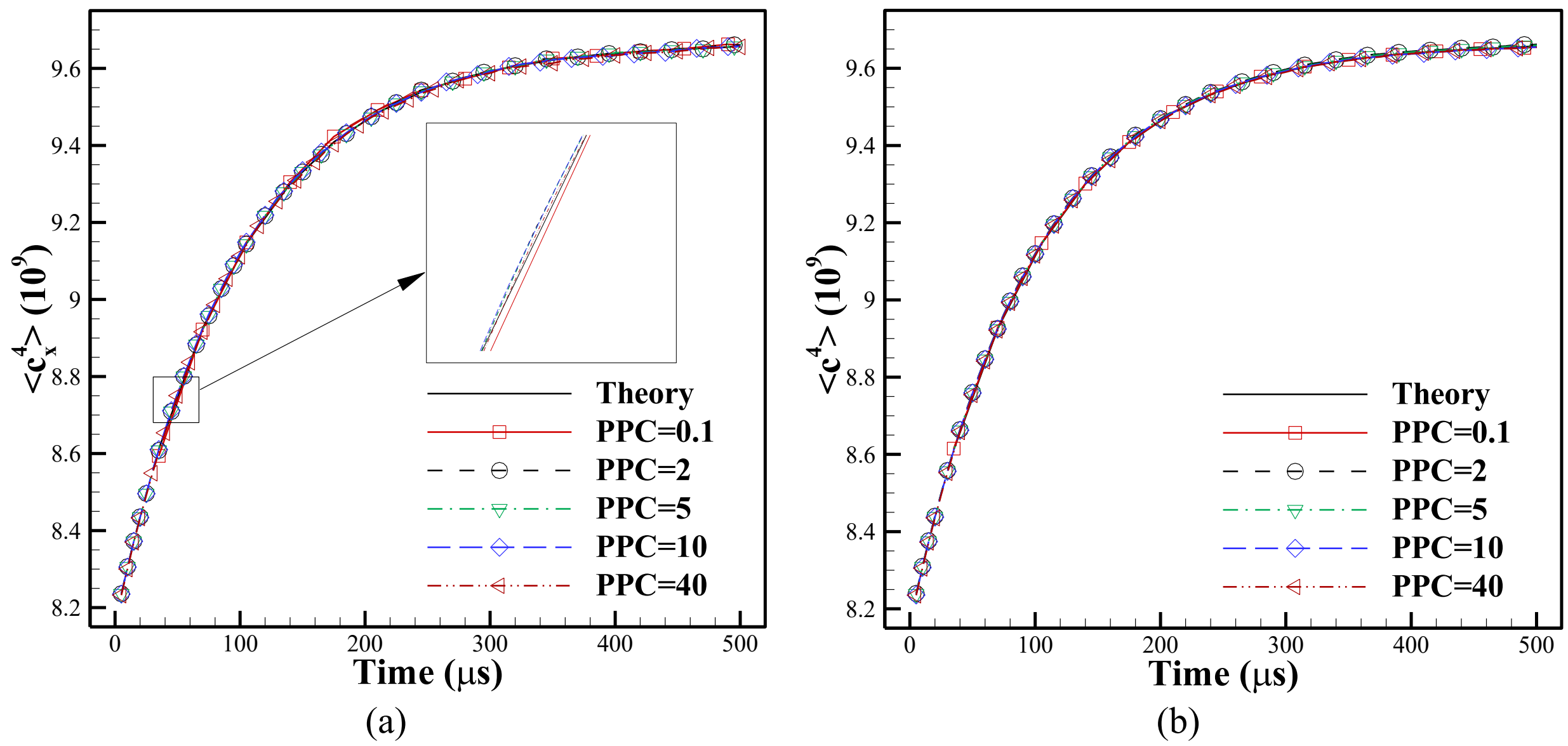

Fig 8. Comparison of the time evolution of the fourth moment of the distribution function of the BKW solution using the SSBT scheme (a) x-velocity component, (b) average of three velocity components.

Figure 9 compares the temporal evolution of the fourth moment of the x-velocity component, as well as the average fourth moment across all three velocity components, obtained using the SGBT scheme for various PPC (PPC=0.1, 5, 10, 40, 200, and 250) and $N_{sel}$, against the corresponding theoretical solution. Figure 9-a indicates the deviation of the results from the theoretical solution for the simulation of the fourth moment of the x-velocity component of the VDF. In contrast, the average of the fourth moment of the three velocity components shows suitable agreement with theory, as shown in Figure 9-b. In our previous paper [41], we described the reason for the deviation in the fourth moment of each velocity component. In that paper, we introduced the concept of the available pair in the cell for each collision scheme. In this regard, the SGBT, like GBT, selects mandatory $N_{sel} < N-1$ pairs and does not have access to all possible pairs of N(N-1)/2 in the cell. In one-dimensional consideration, the particles with small $c_x$ remain longer in the same cell than the other with larger $c_x$. At the same time, $c_y$ and $c_z$ components are arbitrarily large. Thus, a natural separation of particles with small $c_x$ and arbitrary $c_y$ and $c_z$ exists, and such particles stay longer in a given cell. This leads to correlations between the particle's velocity and its effect, which is reflected in the imbalance of the results $\langle c_x^4 \rangle$, $\langle c_y^4 \rangle$ and $\langle c_z^4 \rangle$ as demonstrated, for instance, in Figure 9-c for the simulation using PPC=10 and $N_{sel}$ of N-2 .

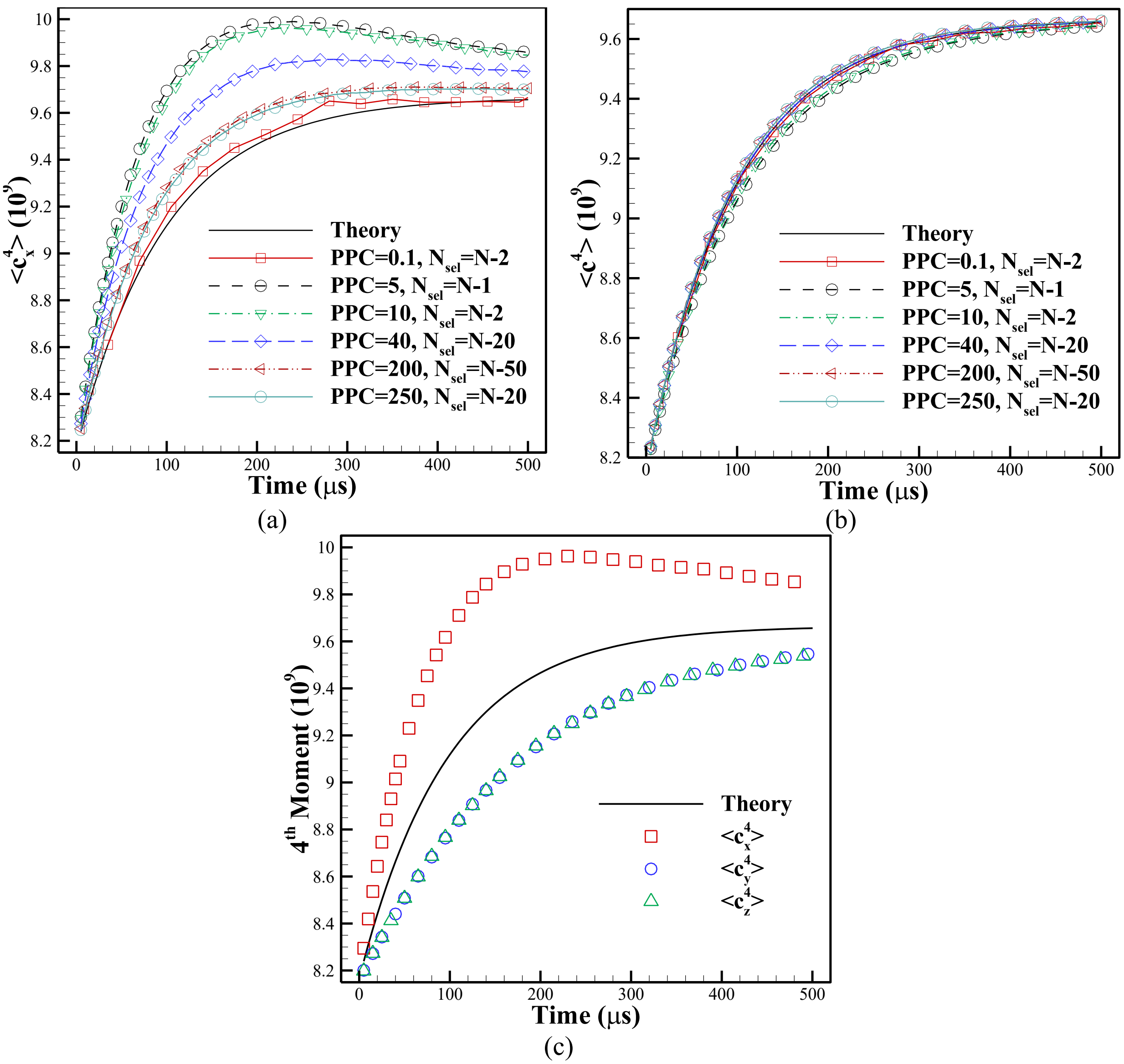

Fig 9. Comparison of the time evolution of the fourth moment of the distribution function of the BKW solution using the SGBT scheme (a) x-velocity component, (b) the average of three velocity components, (c) fourth moment of the component of the VDF.

To address this issue, our previous work proposed simulating this specific relaxation problem on a 3D computational grid. In the current study, we implemented the unsteady relaxation case in a 3D code and conducted simulations using the GBT and SGBT collision schemes to evaluate the effectiveness of this approach. A representative test case, consisting of 10,000 total particles distributed in a single cell with 1,000 subcells, which yields a PPSC (particles per subcell) of 10, as shown in Figure 10, and is compared with the theoretical solution.

Since the performance of the GBT and SGBT schemes is sensitive to the choice of $N_{sel}$, two values, $N_{sel}$=N-3 and N-4, are considered in these simulations. Each frame of this figure compares the fourth moment of the components of the velocity distribution function with the average of 4$^{th}$ moment of the three components of the VDF with theory. As shown, both GBT and SGBT schemes produce results that are in excellent agreement with the theoretical predictions. Moreover, performing the simulation on a 3D computational grid significantly reduces the imbalance in the fourth moments among the velocity components, confirming the validity of the proposed approach.

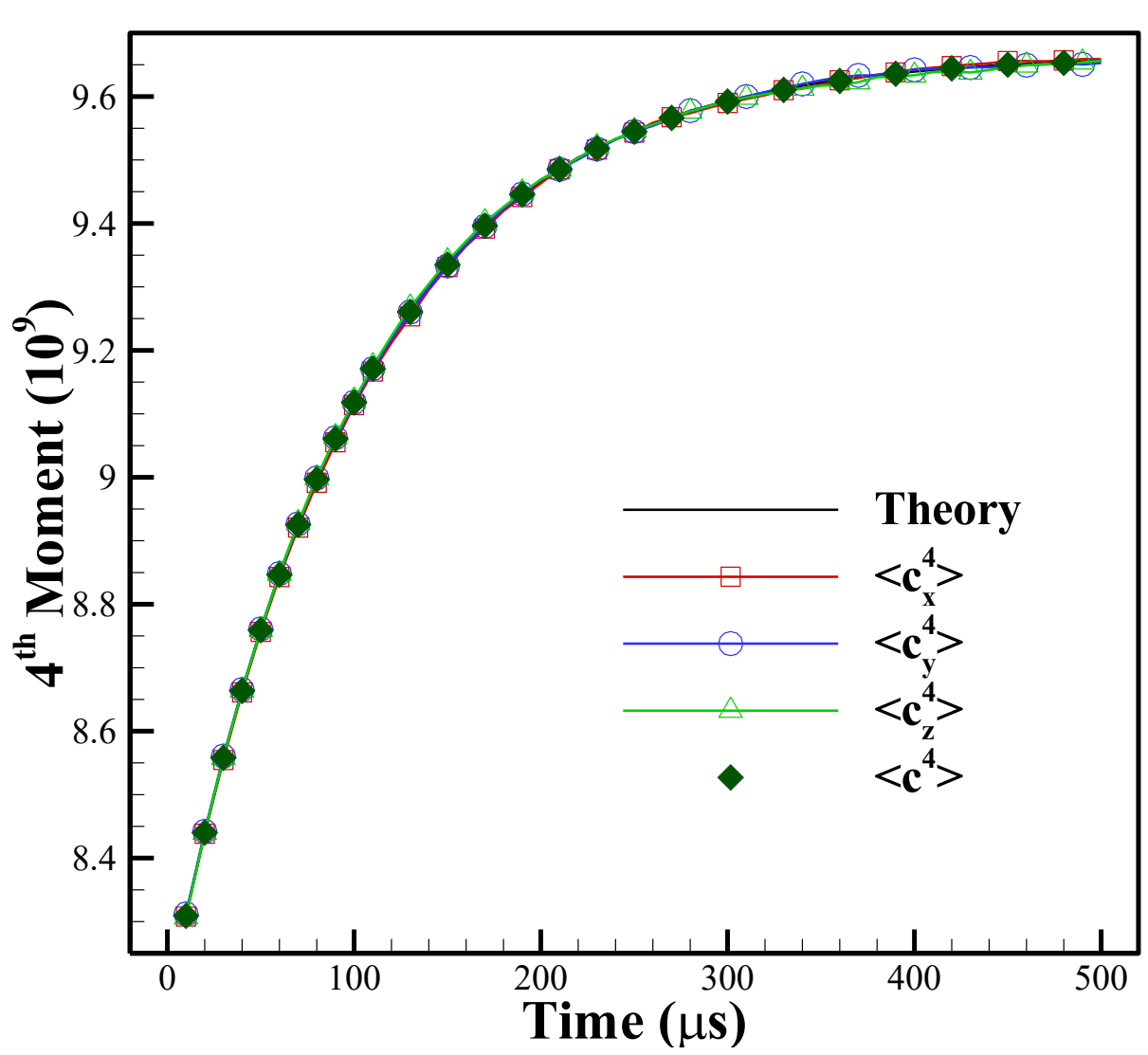

(a)GBT, $N_{sel}$=N-3

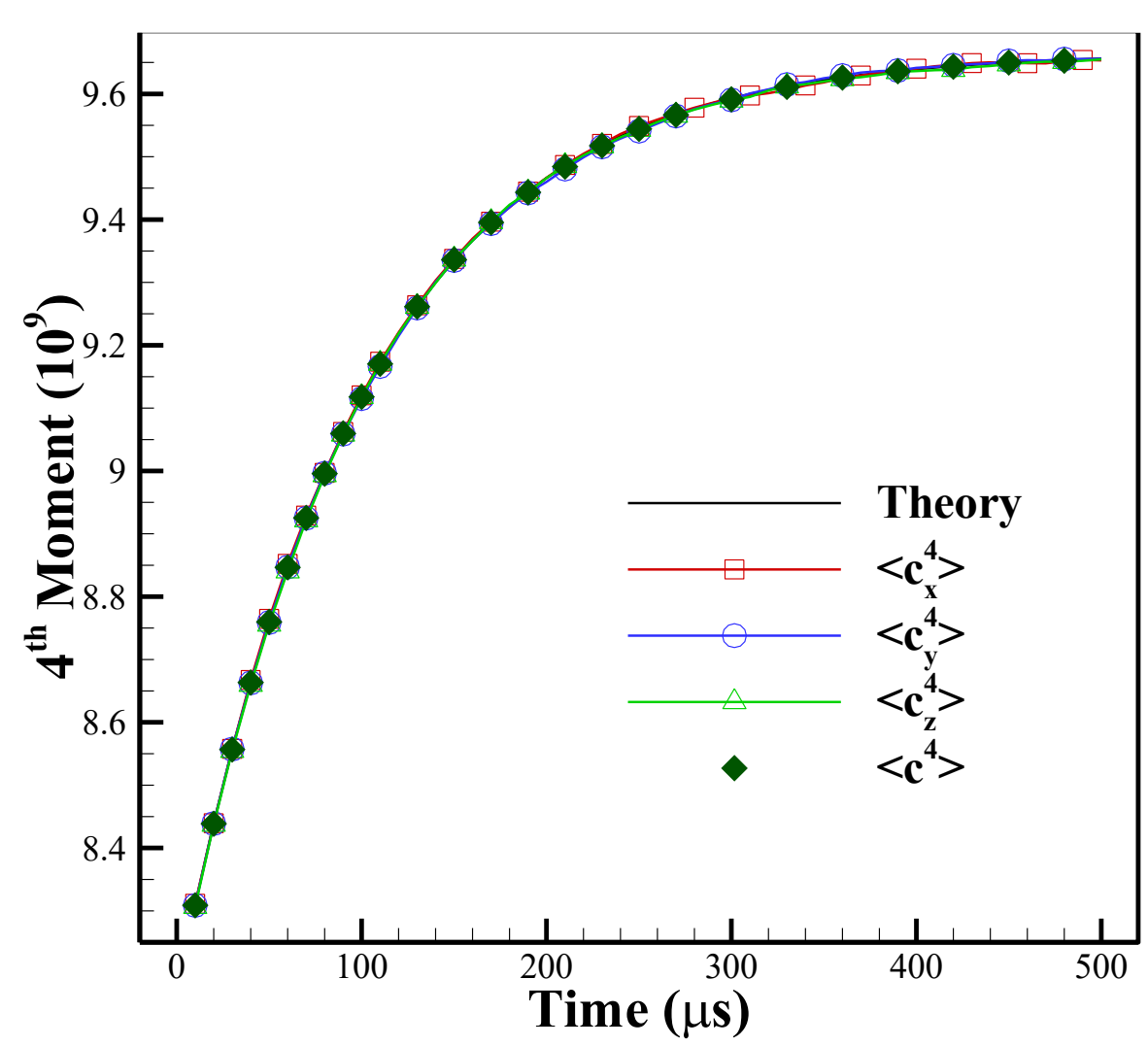

(b)GBT, $N_{sel}$=N-4

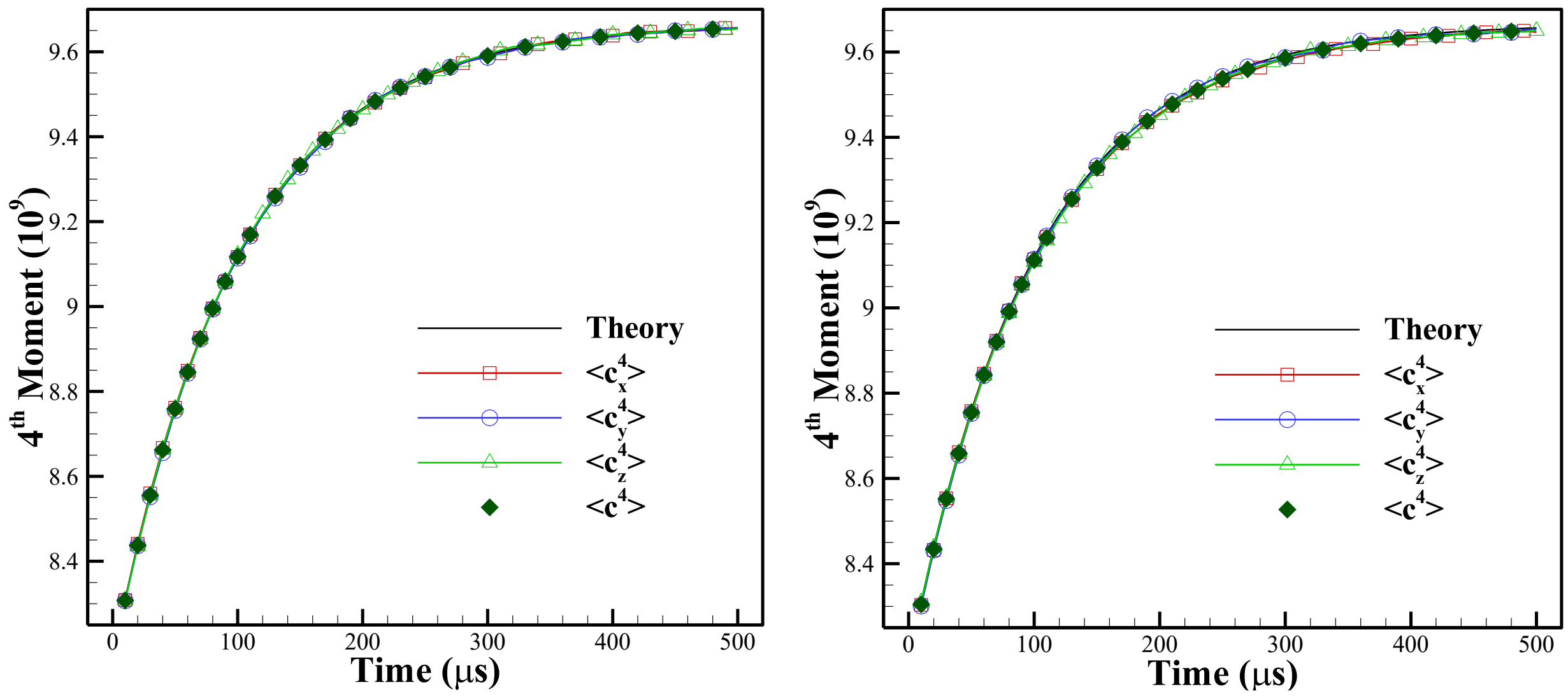

(c)SGBT, $N_{sel}$=N-3 (d)SGBT, $N_{sel}$=N-4

Fig 10. Comparison of the time evolution of the fourth moment of the distribution function of the BKW solution using the different schemes on a 3D grid of subcells (a) GBT, N-3, (b) GBT, N-4, (c) SGBT, N-3, and (d) SGBT, N-4.

### 3.3 Micro-Cavity Flow

The lid-driven microcavity flow is simulated using various Bernoulli Trial-based collision partner selection schemes, including SBT, GBT, SSBT, and SGBT, and the results are compared with those obtained using Bird's standard NTC algorithm. Argon gas is used in the simulations, characterized by a molecular mass of $6.64 \times 10^{-26}$ kg and a molecular diameter of $4.092 \times 10^{-10}$ m. The gas is confined within a square microcavity with a side length $1\times10^{-6}$ m. Figure 11 illustrates the geometry and flow conditions of the test case. The lid of the cavity moves at a constant velocity of 100 m/s, and the Knudsen number is set to 0.01. The cavity walls are maintained at non-uniform temperatures. Specifically, the lower corners of the cavity (points A and D) are set to 350 K. In comparison, the upper corners (points B and C) are maintained at 300 K. This configuration results in a uniform temperature of 300 K along the top (lid) wall and 350 K along the bottom wall.

Meanwhile, the temperatures of the left and right sidewalls vary linearly from 300 K, near the top (lid) to 350 K near the bottom, creating a temperature gradient across the vertical boundaries.

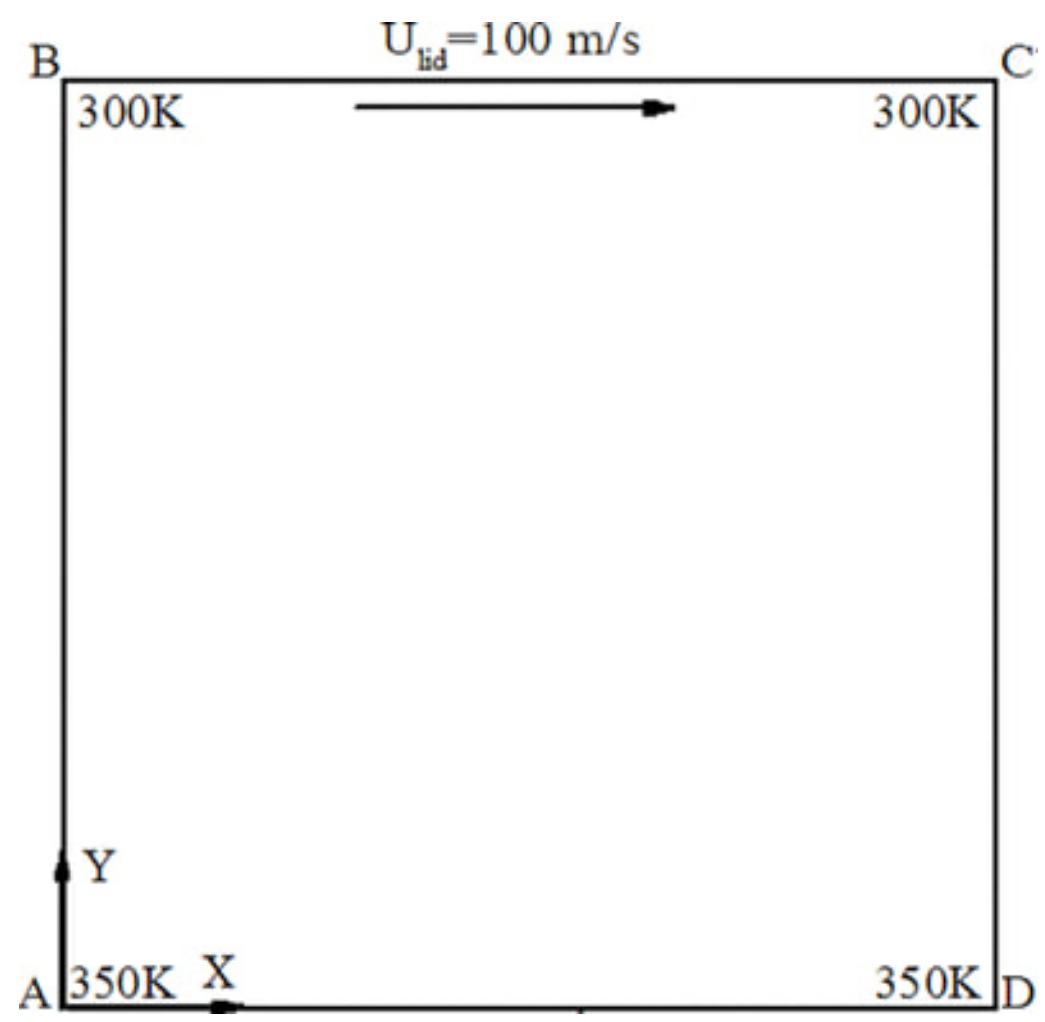

Fig 11. Geometry and boundary conditions of the cavity test case.

### 3.3.1 Performance Analysis

To enable a detailed comparison of the collision partner selection schemes, a performance evaluation is conducted to assess their computational efficiency. While similar comparisons have been carried out previously for the SBT [22], GBT [42], and SSBT [26] algorithms, the primary focus of this section is on the SGBT scheme. In this test, each collision algorithm is used to simulate the lid-driven microcavity flow on a 200 × 200 cell grid with 10 particles per cell (PPC). The simulations begin from an equilibrium condition and proceed until a steady state is achieved.

To determine the point of convergence, the gas temperature near the cavity lid is monitored. A benchmark solution is used as a reference, and the normalized temperature difference between the simulation and benchmark values is computed at each cell along the lid. The simulation is terminated when the sum of these normalized errors across all lid cells falls below a threshold of 0.5. This convergence criterion is mathematically expressed as:

$$Error = \sum_{lidcells} \frac{T_{lid} - T_{Benchmark}}{T_{Benchmark}} \leq 0.5 \quad (23)$$

The benchmark values are obtained from a reference simulation using 50 particles per cell (PPC), conducted over a sufficiently long duration with the standard NTC scheme to ensure convergence. Table 2 presents the results of the performance test. For a comprehensive and rigorous comparison, the table includes key computational performance metrics for the SGBT scheme, alongside corresponding results for the NTC, SBT, GBT, and SSBT schemes.

**Table 2.** Comparative results for cavity flow using different collision partner selection schemes: NTC, SBT, GBT, SSBT, and SGBT.

| | Normalized CPU-time | W_E_R (%) | Col/Sel Ratio (%) | normalized sample size |
|---|---|---|---|---|
| *NTC* | 1 | 0 | 60.58 | 1.00 |
| *SBT* | 1.10 | 3.24 | 35.50 | 1.02 |
| *GBT ($N_{sel}$=N-4)* | 0.62 | 4.99 | 52.71 | 0.64 |
| *SSBT* | 0.68 | 0 | 31.56 | 0.64 |
| *SGBT ($N_{sel}$=N-2)* | 0.80 | 0.02 | 39.44 | 0.78 |
| *N-3* | 0.61 | 0.05 | 45.03 | 0.62 |
| *N-4* | 0.70 | 0.15 | 52.26 | 0.76 |
| *N-5* | 0.68 | 1.28 | 61.68 | 0.75 |
| *N-6* | 0.72 | 0.93 | 48.56 | 0.87 |
| *N-7* | 1.22 | 0.91 | 42.45 | 1.60 |
| *N-8* | 1.06 | 1.08 | 40.32 | 1.48 |

The first column of Table 2 reports the CPU time required to reach the stopping criterion, with all values normalized to the CPU time of the NTC scheme. The results show that, apart from the cases where $N_{sel}$=*N-7* and *N*-8, the SGBT scheme consistently achieves faster performance compared to NTC. This is because the number of selected pairs is reduced. At the same time, the time step constraints for achieving a low exceedance ratio do not compromise the efficiency of the scheme. It is important to note that, for the BT-based schemes to yield physically accurate results, the proportion of collisions with a probability exceeding one, referred to as the W exceed ratio (W_E_R), must be constrained to approximately 1%. To satisfy this condition, all simulations listed in Table 2 were performed using a uniform time step of $1.63\times10^{-11}$ s, except for *N-7* and *N-8* cases of the SGBT scheme. For these two configurations, the time step was adjusted to ensure that the W_E_R remained below the 1% threshold, increasing computational time. As indicated in the table, the SSBT scheme yields shorter computation times compared to the SBT method. Furthermore, the GBT with $N_{sel}=N-4$ and the SGBT using $N_{sel}=N-3$ complete the simulation in slightly less time than the SSBT. Both GBT and SGBT also require less time to reach the stopping condition than the SBT and NTC schemes, except in cases involving N-7 and N-8.

Figure ۱۲-a presents the collision probability coefficient for the GBT and SGBT schemes, while Figure ۱۲-b displays the same coefficient for the SBT and SSBT schemes. These plots correspond to a cell with 25 particles per cell (PPC), where the number of selected partners, $N_{sel}$, is set to 20. For the GBT scheme, where the probability coefficient depends on the order of the selected particle pair in the list, the coefficient was calculated for the second particle in the list ($i$=2). In contrast, the coefficients shown in Figure ۱۲-b were computed for every particle in the cell. The results from both figures indicate that, under identical time step and velocity conditions, the SBT and GBT schemes exhibit higher probability coefficients than the SSBT and SGBT methods. This implies a higher likelihood of exceeding the upper bound of the collision probability (i.e., values greater than 1) in the SBT and GBT schemes. These observations are consistent with the results in Column 2 of Table 2 and Figure ۱۲-c, where the W_E_R (probability exceedance ratio) values for GBT and SGBT are shown without adjusting the time step to constrain W_E_R near 1%. As illustrated, the higher probability coefficients in the GBT scheme lead to larger W_E_R values across all $N_{sel}$ settings compared to those in the SGBT scheme.

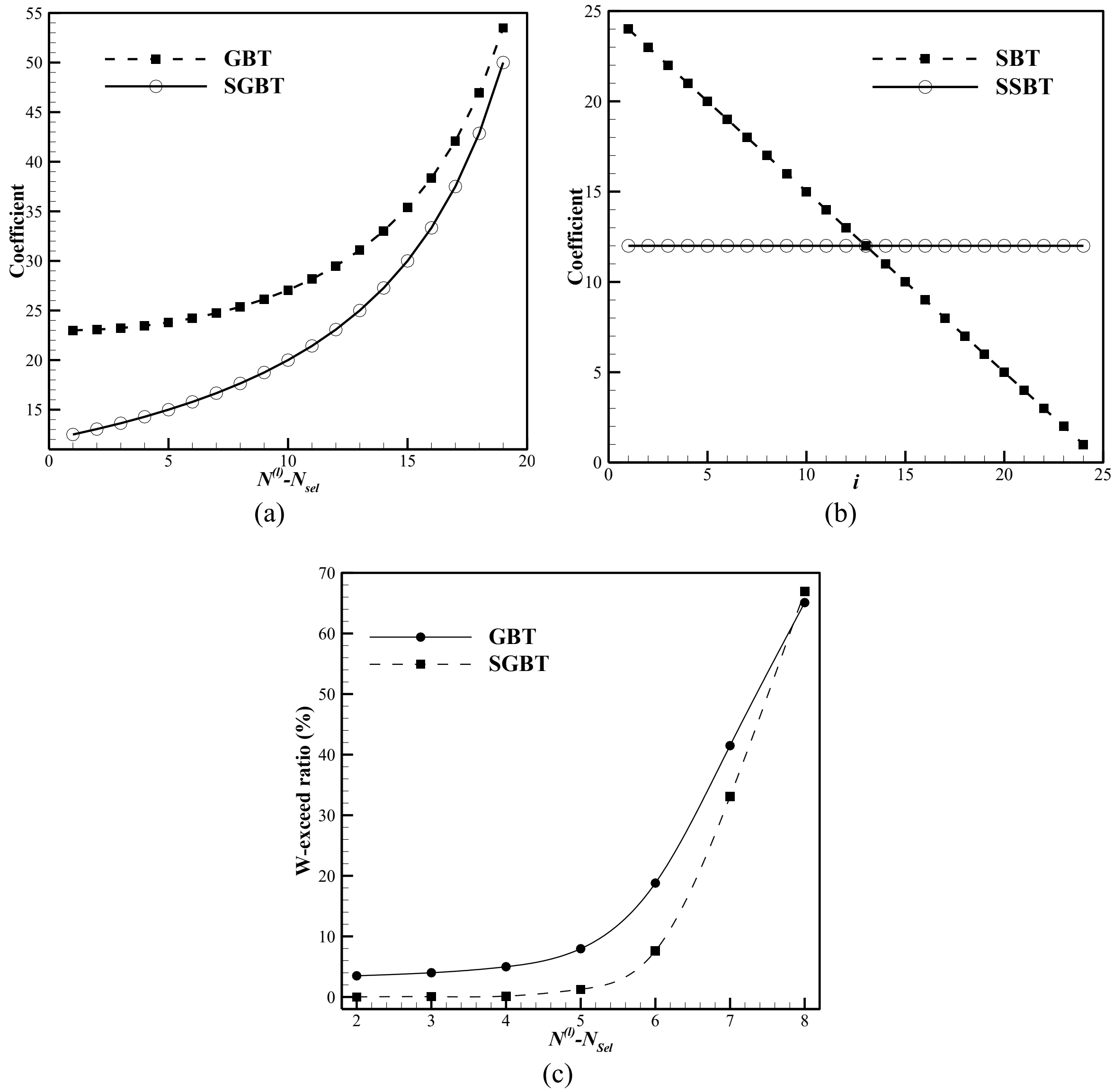

Fig ١٢. Comparison of collision probability coefficients and exceed ratios: (a) GBT vs. SGBT, (b) SBT vs. SSBT, and (c) comparison of *W* exceed ratios between GBT and SGBT schemes.

The third column of Table 2 shows the ratio of accepted collisions to the total number of selected particle pairs. For the SGBT scheme, this ratio increases as $N_{sel}$ decreases from *N-2* to *N-8*. This trend can be attributed to the differences between the SSBT and SGBT schemes. In the SSBT approach, due to the use of a relatively small time step, required to ensure that the collision probability does not exceed unity, a significant portion of selected particle pairs are ultimately rejected, as they do not meet the collision criterion. To address this limitation, the SGBT scheme, similar to the GBT, was designed to reduce the number of selected pairs while increasing the likelihood that a selected pair will undergo a collision. Unlike GBT, however, the SGBT method employs a more uniform collision probability distribution, resulting in a lower probability exceedance ratio (W_E_R), as shown in Figure ١٢-c. As mentioned earlier, for each value of $N_{sel}$ in Table 2, the time step is adjusted to ensure that the W_E_R remains below 1%. This adjustment accounts for the slight reduction in the collision-to-selection ratio observed for N-6, N-7, and N-8 compared to N-5.

The final column of Table 2 presents the sample size required for each scheme to reach the predefined stop criterion, with values normalized relative to the sample size of the NTC scheme. This metric provides clearer insight into the computational effort needed by each collision algorithm to achieve convergence. Among the SGBT configurations, the most efficient performance is observed when $N_{sel}$=N-3, where the sample size is reduced by approximately 38% compared to the NTC scheme and by 40% compared to the SBT. For this selection parameter, the required sample size is comparable to that obtained using the GBT with $N_{sel}$=N-4 and the SSBT scheme.

### 3.3.2 Flow field properties

To provide a comprehensive evaluation of the different collision partner selection schemes, flow field properties are examined across a range of Knudsen numbers. The simulations are carried out using several schemes, including NTC, SBT, GBT, SSBT, and SGBT. The Knudsen number varies from 0.01, corresponding to the slip flow regime, up to 10, representing the upper limit of the transition regime. Each simulation uses a grid of 200×200 cells with an initial particles-per-cell (PPC) value of 10. For the GBT and SGBT schemes, the number of selected particles was set to $N_{sel}=N-4$.

Figure ۱۳ shows the velocity magnitude contours obtained using different collision partner selection schemes at various Knudsen numbers. In each frame, which represents the velocity magnitude contours for a specific Knudsen number, the background is filled with the contours from the NTC scheme. To highlight the results from other collision schemes, the flow domain is divided into columns, each corresponding to a different scheme. From left to right, the columns display the SBT (dashed line), GBT (dash-dot line), SSBT (dash-dot-dot line), and SGBT (solid line) schemes, with the NTC contours in the background. This layout enables easy comparison of the different BT schemes against the NTC in various parts of the domain, as well as between the schemes at column boundaries. The figure demonstrates a good level of agreement between all schemes for simulating the velocity magnitude contours.

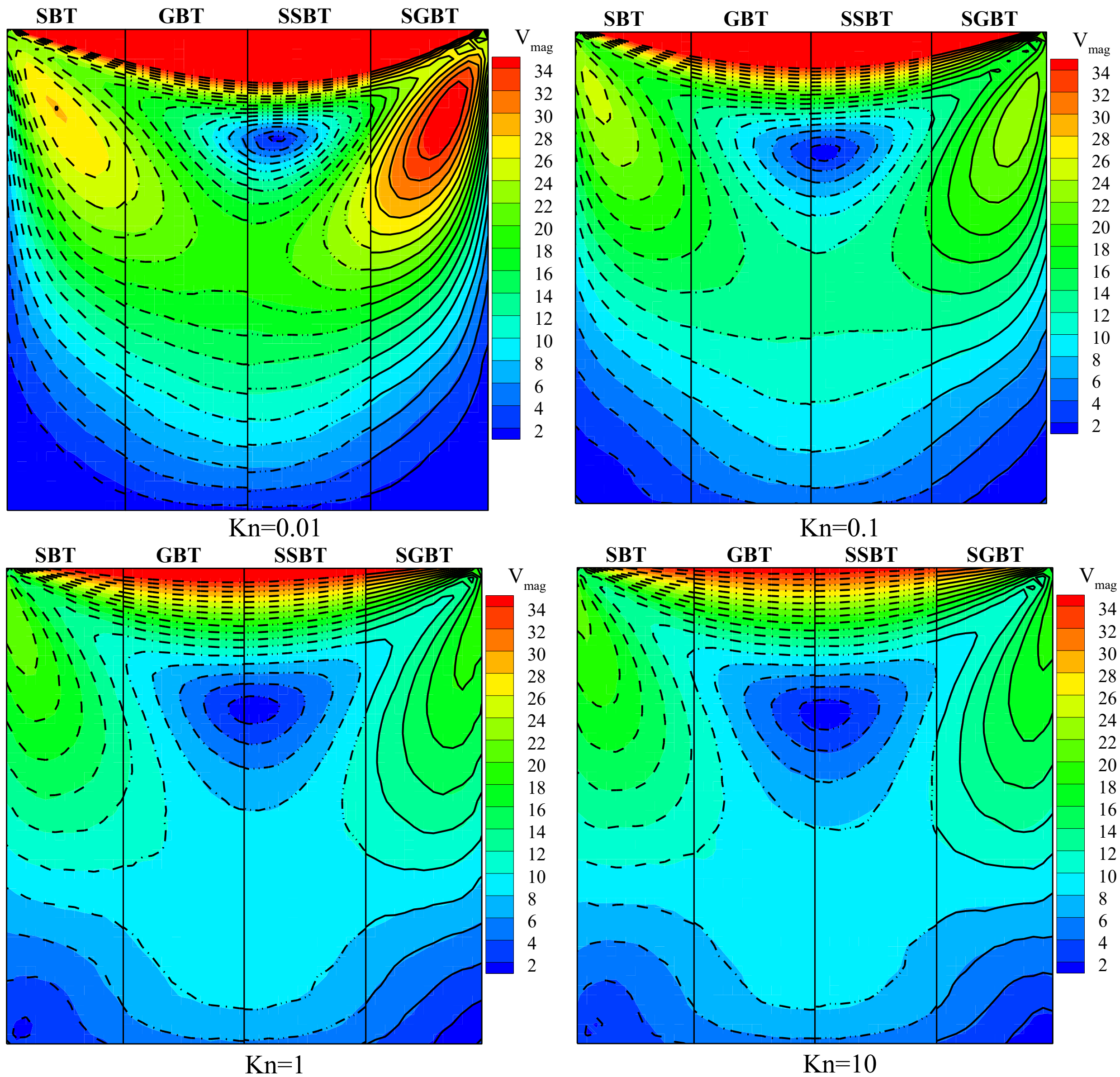

Fig ۱۳. Velocity magnitude contours at various Knudsen numbers: comparison of SBT (dashed line), GBT (dash-dot line), SSBT (dash-dot-dot line), and SGBT (solid line) schemes overlaid on the NTC results (background flood contours).

To further investigate the ability of the particle pair selection schemes to simulate flow field properties, we present the temperature contours from the simulations at different rarefaction regimes in Figure ۱۴. The flood and lines are indicated in the same manner as in Figure ۱۳. Figure ۱۴ shows excellent agreement between the temperature fields calculated by all the schemes.

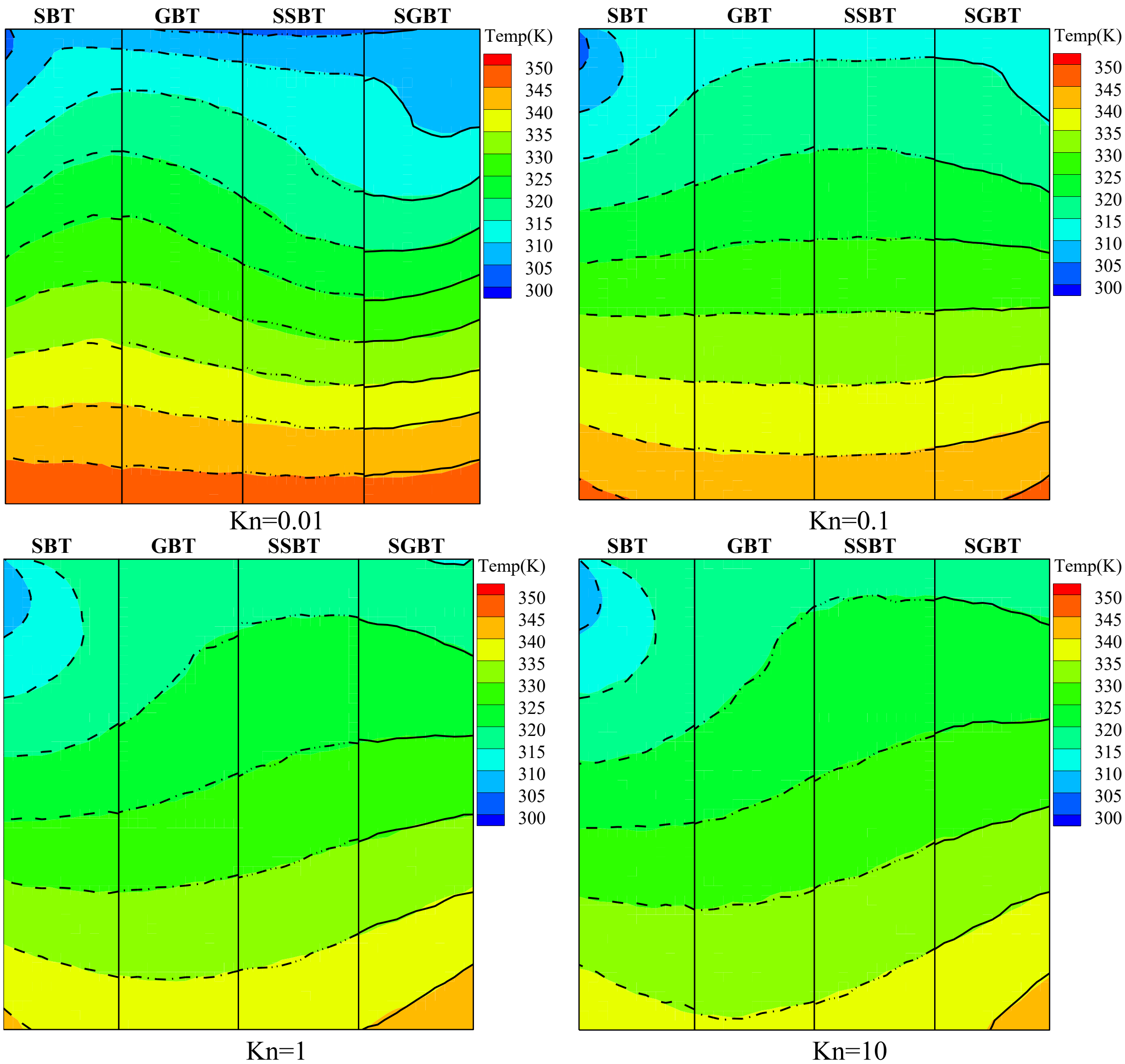

Fig ۱۴. Temperature contours at various Knudsen numbers: comparison of SBT (dashed line), GBT (dash-dot line), SSBT (dash-dot-dot line), and SGBT (solid line) schemes overlaid on the NTC results (background flood contours).

### 3.3.3 Circulation and the location of the vortex center

To examine the nonequilibrium effects on vortex strength within the microcavity, the circulation of the mean velocity field, denoted as, $\Gamma$ is defined as follows.

$$\Gamma = \oint V.ds = \int rot_n V.dA = \sum_{i,j} \left[ \frac{(V_{yi+1} - V_{yi,j})}{\Delta x} - \frac{(V_{xj+1} - V_{xi,j})}{\Delta y} \right] \Delta x \Delta y \tag{24}$$

The summation is carried out over all computational cells within the cavity. Figure 1۵-a illustrates the variation of flow circulation with the Knudsen number in the driven cavity, based on simulations using different collision partner selection schemes. As shown, all schemes yield nearly identical circulation magnitudes. In this figure, the circulation values are normalized using $\Gamma_0 = U_{wall} \times L$. From a physical point of view, increasing nonequilibrium effects progressively weaken the shear forces induced by the moving lid, thereby reducing the vortex strength within the cavity. Moreover, this reduction trend becomes significantly less pronounced as the Knudsen number exceeds 0.5.

Figure 1۵-b illustrates how the location of the vortex center varies with the Knudsen number. As the Knudsen number increases, the corresponding Reynolds number decreases, leading to enhanced viscous dissipation within the flow field. This increased dissipation causes the kinetic energy introduced by the moving lid to decay more rapidly, shifting the vortex center leftward toward the geometric center of the cavity. Additionally, with increasing rarefaction, the vortex center also moves downward, away from the driven lid and closer to the bottom wall. This downward shift is likewise a result of the reduced Reynolds number at higher Knudsen numbers. From a numerical perspective, the figure shows that all collision partner selection schemes produce nearly identical predictions for the vortex center location.

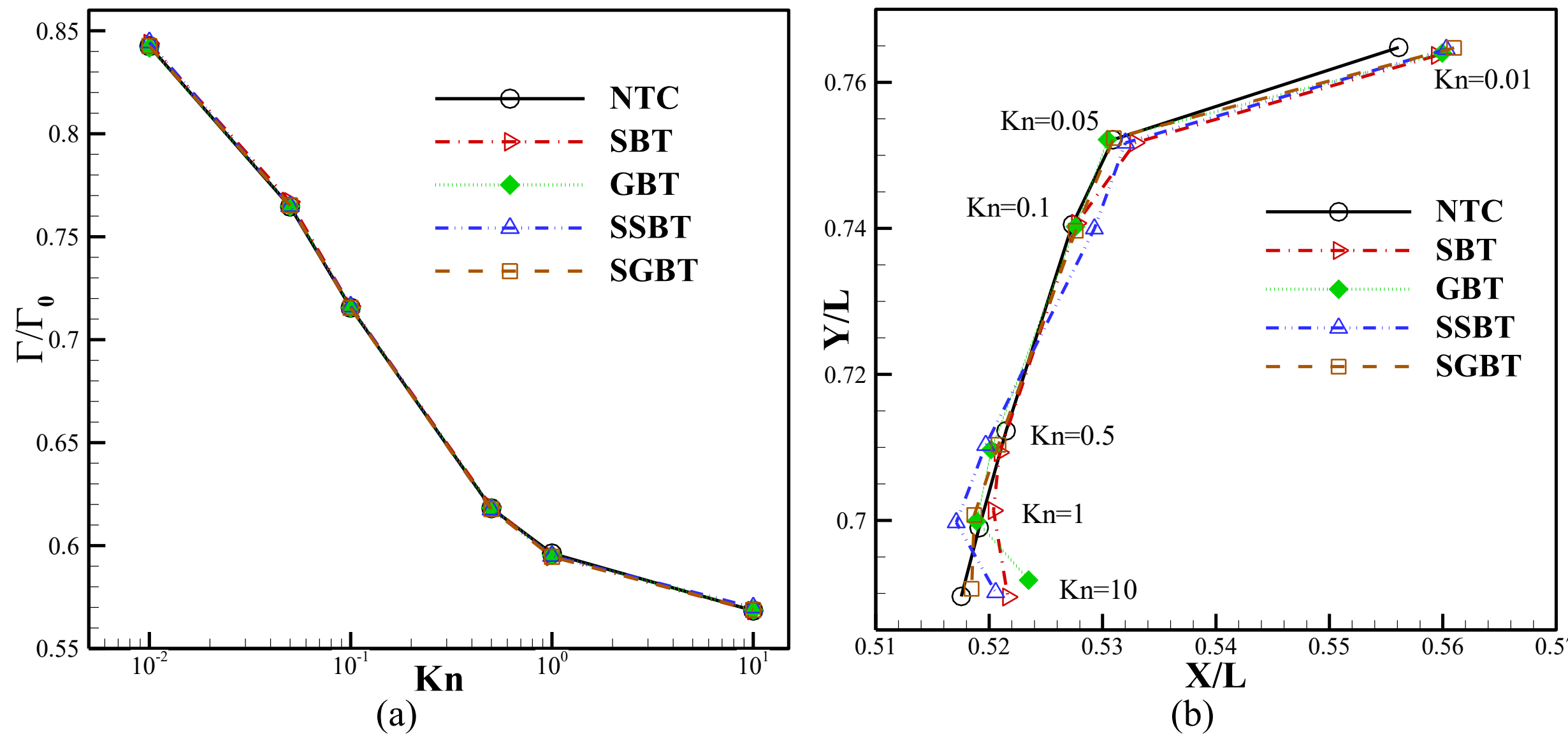

Fig 1۵. (a) Normalized circulation of the velocity field, and (b) position of the vortex center within the cavity, as a function of the Knudsen number for different collision partner selection schemes.

### 3.3.4 Velocity slip and temperature profiles

Figures 1۶-a and 1۶-b display the normalized horizontal and vertical components of the velocity vectors along the vertical and horizontal centerlines of the cavity, respectively. These plots reveal a clear trend: as the Knudsen number increases, the curvature of the horizontal velocity profile decreases. This behavior directly correlates with the reduction in vortex strength observed at higher Knudsen numbers, as previously shown in Figure 1۵-a. The weaker circulation results in a more uniform velocity distribution, especially near the core of the cavity, reflecting the diminished influence of the driven lid due to rarefaction effects. Figures 1۶-c and 1۶-d present the velocity magnitude and temperature profiles along the horizontal centerline of the cavity. These results further confirm that all collision partner selection schemes yield nearly identical predictions across the entire range of Knudsen numbers studied. The consistency of these profiles demonstrates that each scheme is equally capable of capturing the essential flow features and thermal behavior,

regardless of the degree of rarefaction. This suggests that the accuracy of the different collision partner selection approaches remains robust even as the flow transitions from near-continuum to highly rarefied regimes, underscoring their reliability for simulating microscale gas flows.

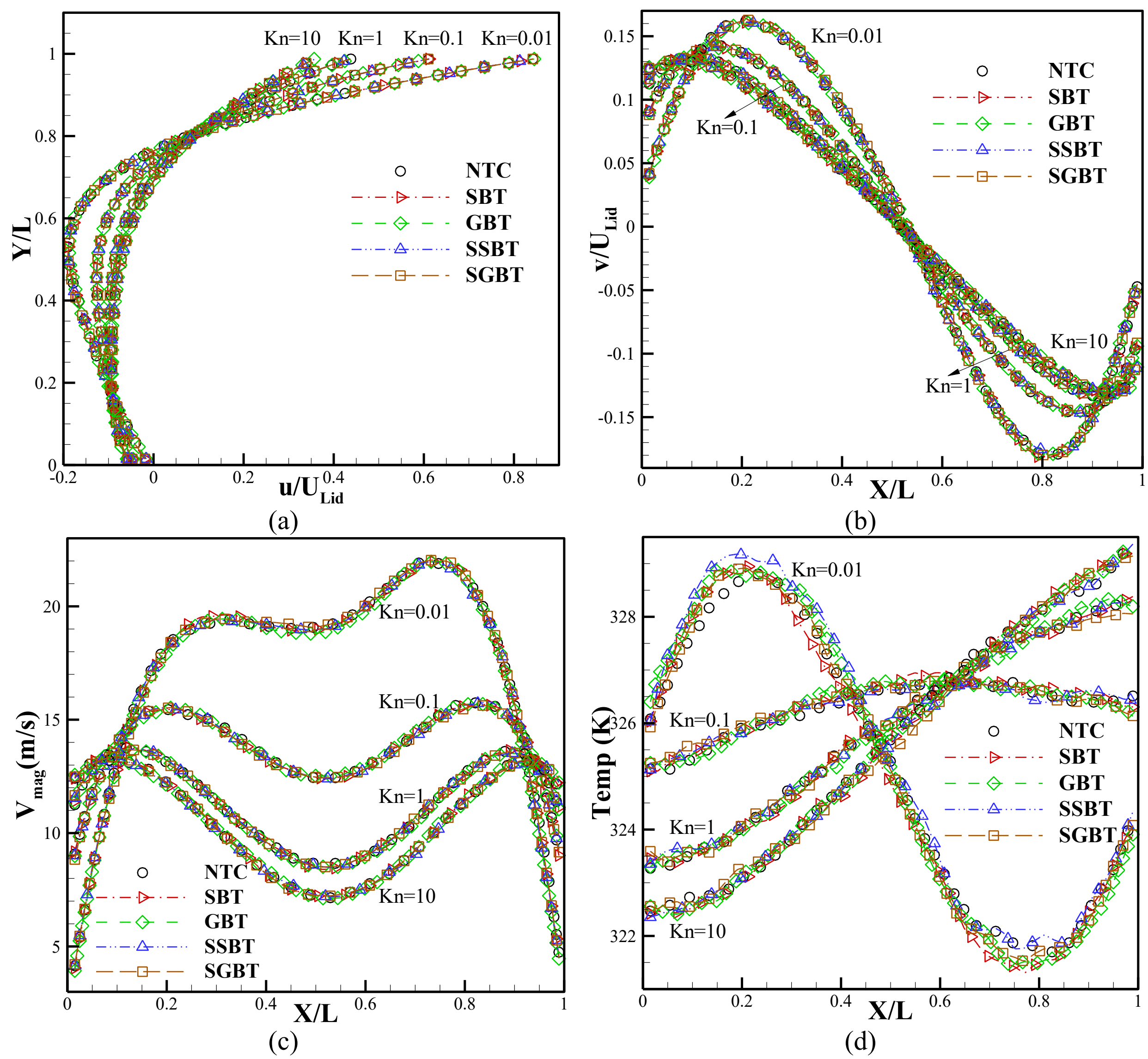

Fig 1۶. Flow properties along the cavity centerlines obtained using different collision partner selection schemes, (a) horizontal velocity component, *u*, along the vertical centerline, (b) vertical velocity component, *v*, along the horizontal centerline, (c) and (d) velocity magnitude and temperature distribution along the horizontal centerline

Figure 1۷-a and 1۷-b present the dimensionless velocity slip and temperature jump along the surface of the driven lid, respectively. Both plots demonstrate excellent agreement among the results obtained from the NTC, SBT, GBT, SSBT, and SGBT collision partner selection schemes. The close alignment of these results confirms that all schemes are equally capable of accurately capturing the non-equilibrium boundary effects. In particular, the velocity slip observed in Figure 1۷-a reflects the deviation from the classical no-slip boundary condition due to rarefaction. In contrast, the temperature jump in Figure 1۷-b indicates the discontinuity in temperature at the gas-solid interface. The ability of all schemes to consistently predict these effects across the range of Knudsen numbers studied highlights their robustness and reliability in modeling microscale flows, where boundary interactions play a critical role in determining overall flow behavior.

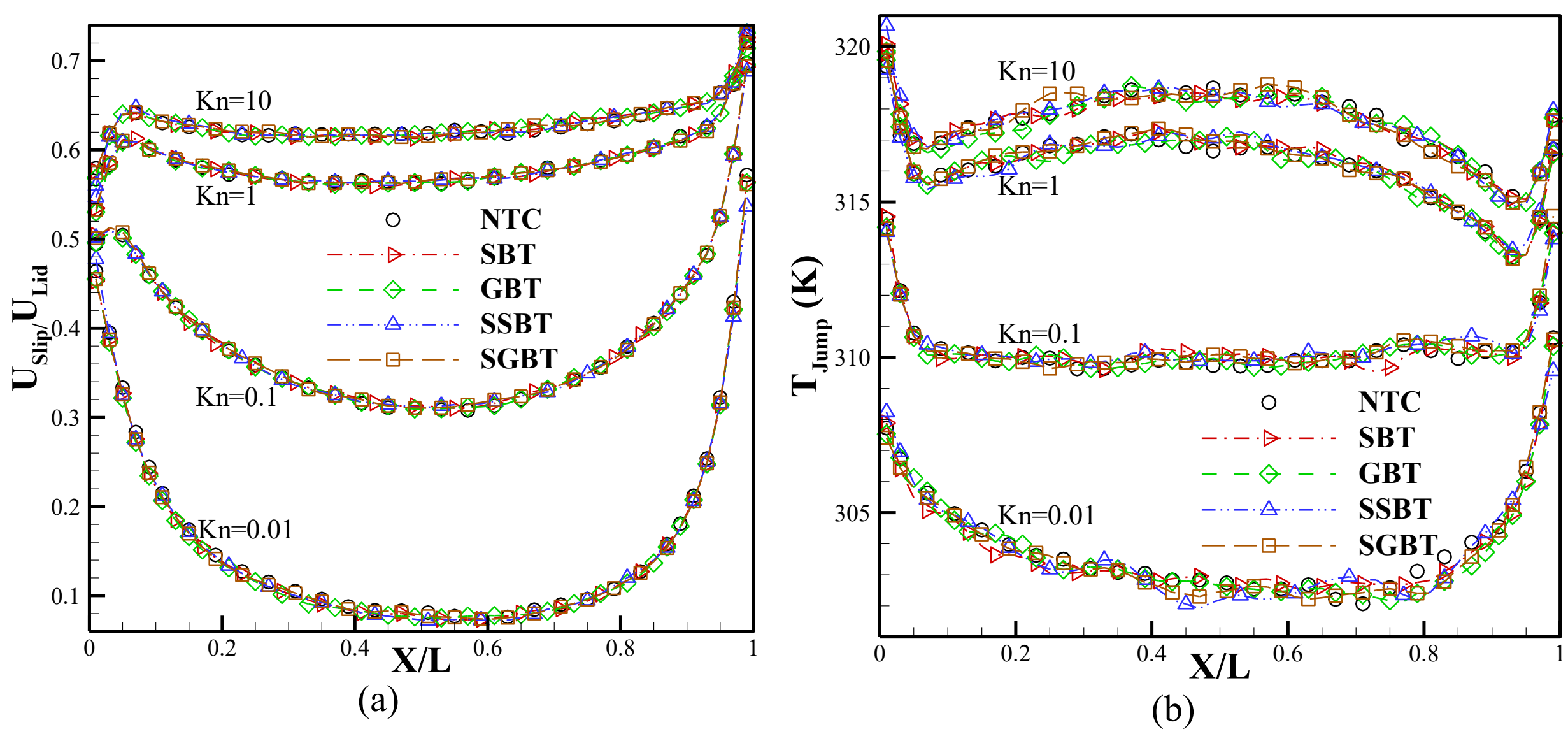

Fig 1۷. (a) Normalized velocity slip and (b) temperature jump along the driven lid.

### 3.4 Hypersonic Flow over a Circular Cylinder

To evaluate the performance of various collision schemes, a benchmark simulation of rarefied hypersonic flow over a circular cylinder was performed. Researchers frequently use this test case, i.e., [32,35,42-43], because it presents a wide range of collision frequencies throughout the domain, from the high-frequency region near the upstream stagnation point to the low-frequency region behind the cylinder, these variations make it a challenging and comprehensive scenario for assessing the accuracy and robustness of collision algorithms. In this simulation, rarefied argon gas flows at Mach 10 (U=2634.1 m/s) with a freestream temperature of 200 K over a 12-inch diameter cylinder. The cylinder surface is fully diffusive and maintained at a high temperature of 6500 K.

Figure 1٨ illustrates the computational domain and the boundary conditions applied. Given that hypersonic flow over a cylinder involves significant gradients in collision frequency and mean free path, an adaptive subcell grid is necessary to accurately capture these variations [35]. To address this, we implemented BT-based collision algorithms combined with the transient adaptive subcell (TAS) technique within the DS2V code developed by G. Bird, as detailed in his latest monograph [38]. For all simulations presented here, the reference solution was obtained using the SBT-TAS algorithm with an adequately high number of particles per cell (PPC) and a refined grid. The grid independence and PPC sensitivity tests for this case were conducted in our previous study [42], where an optimal grid resolution of 194×100 divisions and 11.5 particles per cell was determined.

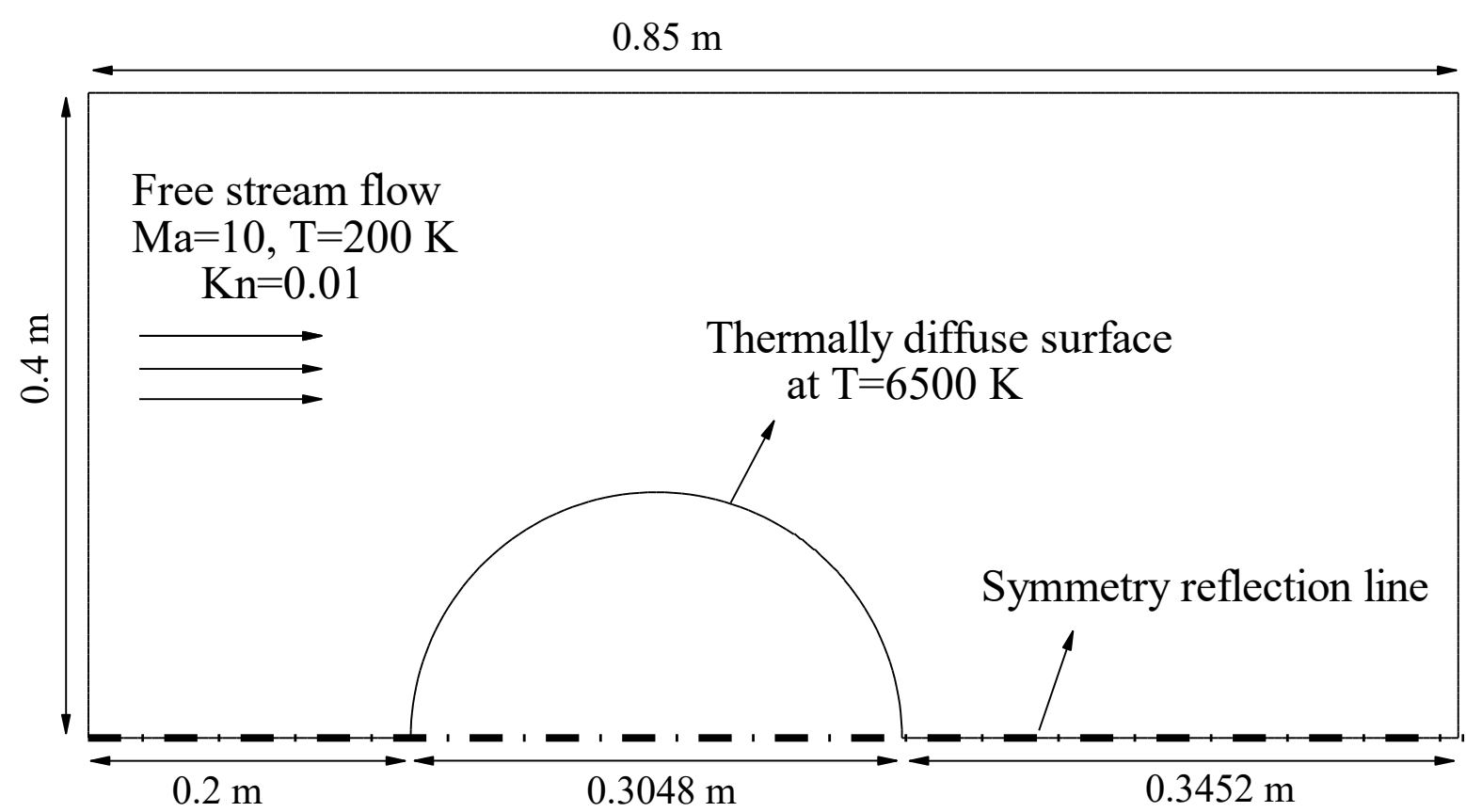

Fig 1۸. Schematic of the computational domain and boundary conditions for the two-dimensional hypersonic flow over a circular cylinder.

### 3.4.1 Flow field properties evaluation

This section evaluates the capability of different collision partner selection algorithms to reproduce flow field properties. While the performance of the SBT, GBT, and SSBT schemes has been addressed in our previous studies [26,42-43], the focus here is on presenting the results obtained using the SGBT algorithm. Figure 1۹ displays temperature contours generated by the SGBT scheme for various $N_{sel}$ values (represented by solid lines), overlaid on the reference solution obtained from the SBT-TAS method (shown as the background color field). As shown, the SGBT results for all examined $N_{sel}$ values exhibit good agreement with the reference solution, indicating the method's accuracy and robustness in resolving the flow field properties.

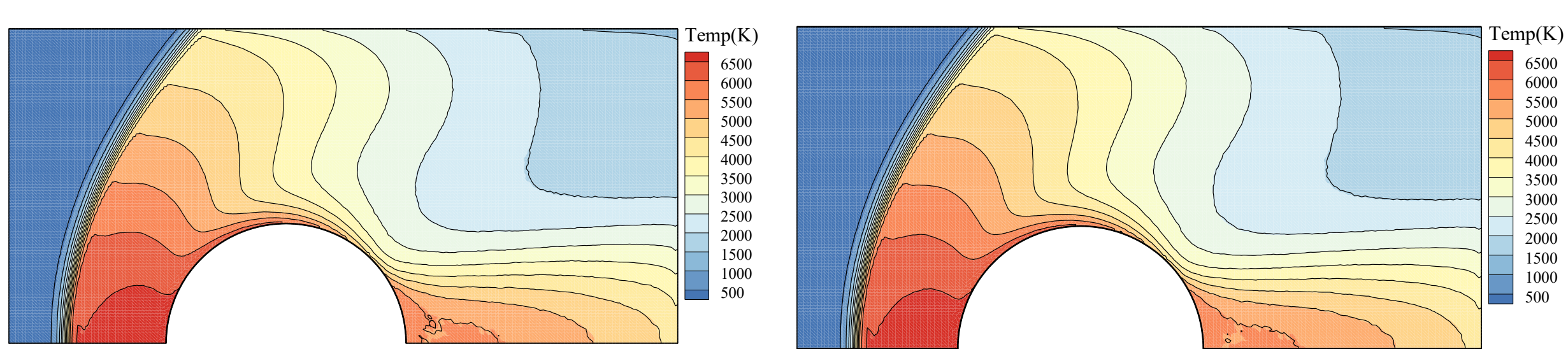

$N_{sel}$=$N$-2 $N_{sel}$=$N$-3

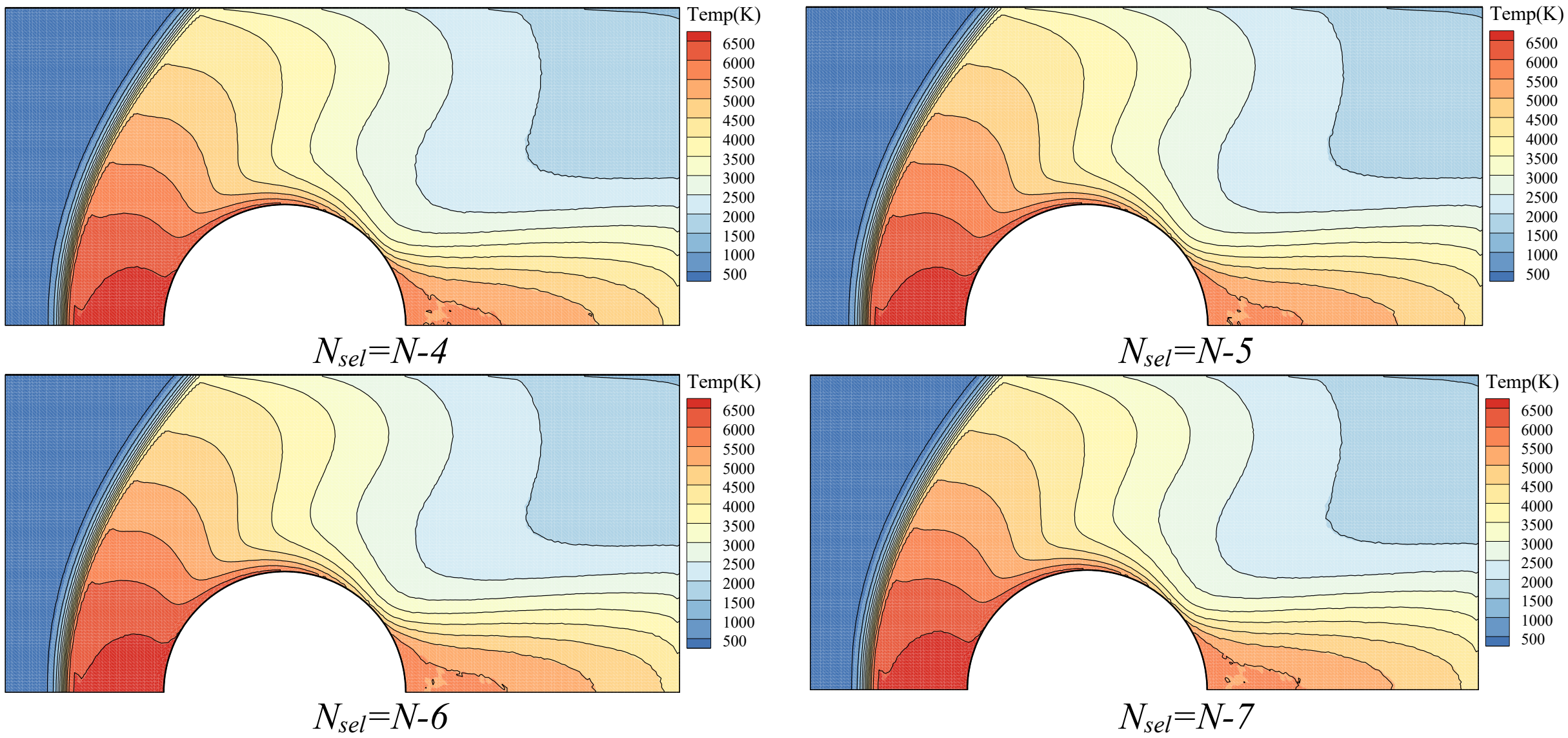

Fig 1٩. Temperature contours (SBT-TAS: flood plot, SGBT-TAS: contour plot).

To facilitate a comprehensive comparison of collision partner selection schemes, Figure ٢٠ shows the velocity and temperature contours obtained from the NTC, SBT, GBT, SSBT, and SGBT methods. Solid lines represent the results of each scheme, overlaid on the background contours of the NN solution. The GBT and SGBT schemes were simulated using $N_{sel}$=N-4. Velocity magnitude contours are presented in frame (a), while temperature contours are shown in frame (b). Both frames demonstrate excellent agreement between the results obtained from various schemes and the NN solution.

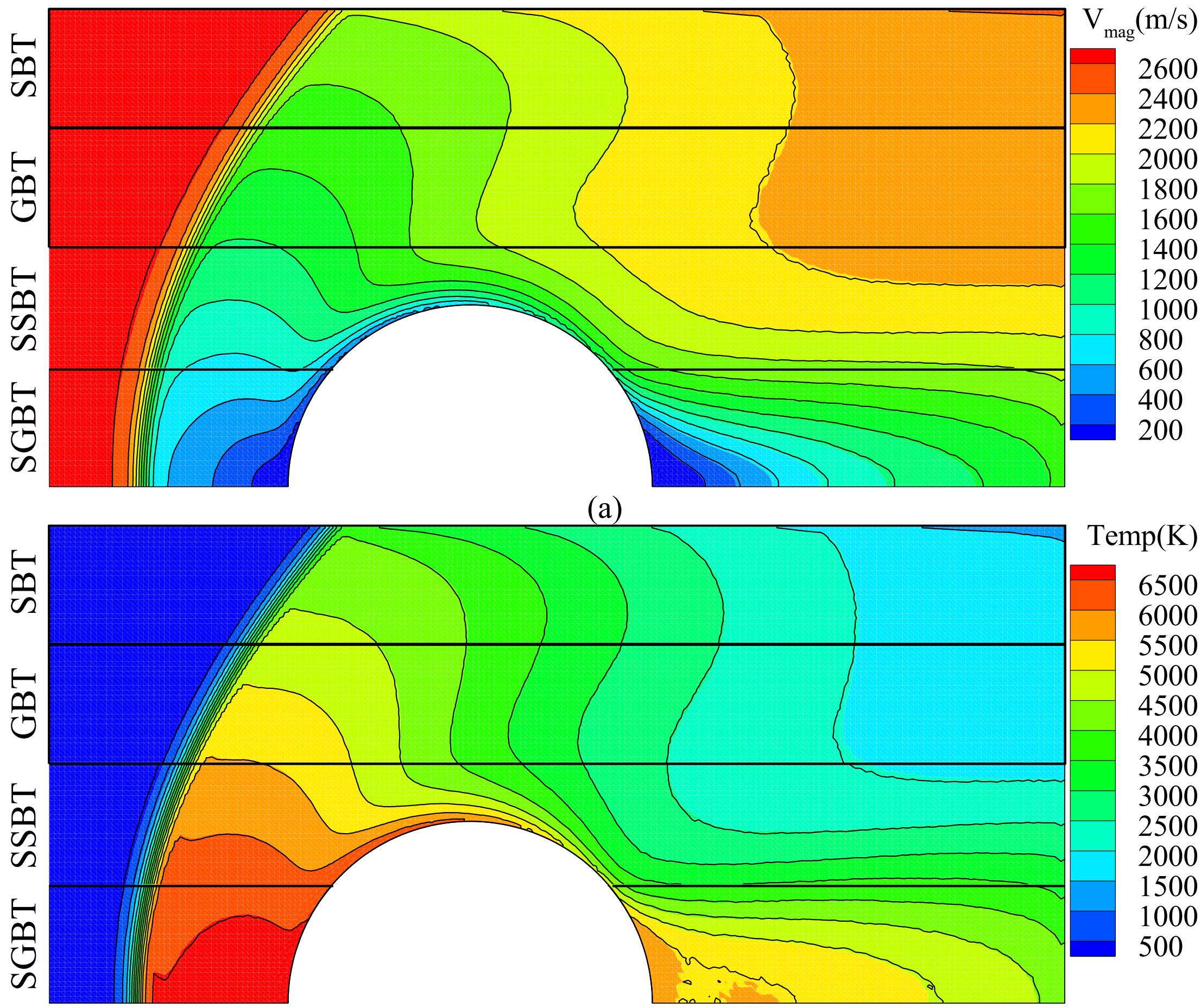

(b)

Fig ٢٠. Comparison of flow field properties from different schemes (a) velocity, (b) temperature, (Flood: NN, Lines: BT-base schemes)

Another significant characteristic of hypersonic flow is the formation of a shock wave. Figure ٢١ compares flow properties along the stagnation line in front of the cylinder, obtained from different computational schemes, with the SBT-TAS solution serving as the reference. Figure ٢١-a shows the Mach number and temperature profiles, both normalized by their respective free-stream values. The SSBT, GBT, and SGBT schemes demonstrate good agreement with the reference results. At the same time, the NN solution exhibits a slightly thicker shock layer in the Mach number profile, indicating a modest discrepancy in shock capturing. As the figure shows, the GBT and SGBT have been performed using $N_{sel}$=N-4. The normalized density distribution, presented in Fig. ٢١-b, shows excellent agreement across all methods compared to the SBT-TAS solution. These results confirm the reliability of the tested schemes in resolving shock structures in hypersonic regimes.

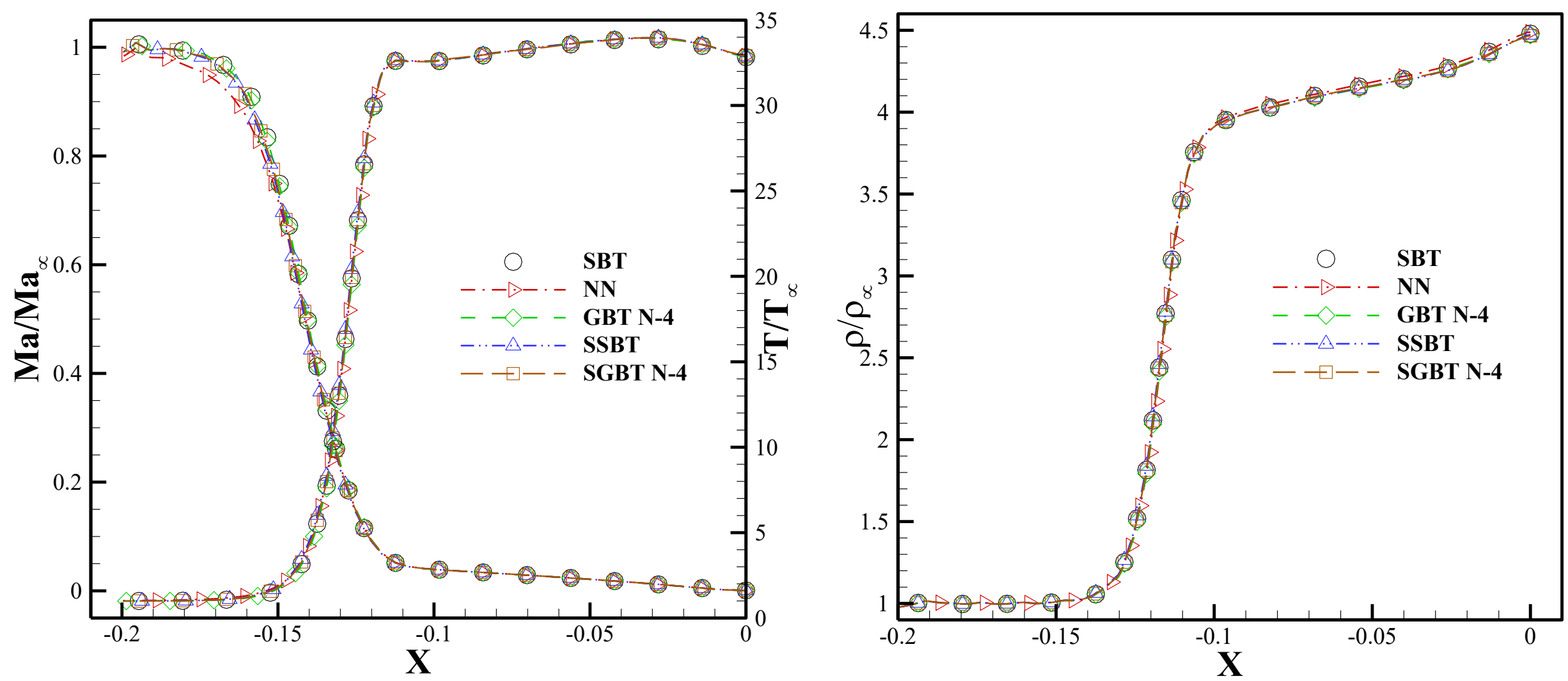

(a) (b)

Fig ٢١. (a) Normalized Mach number and temperature profiles; (b) normalized density profile along the stagnation line in front of the cylinder.

### 3.4.2 Surface properties evaluation

In this subsection, the simulation results for velocity, temperature, heat flux, and pressure coefficient on the cylinder surface have been reported. The comparison of velocity and pressure coefficient distributions on the cylinder surface, as shown in Fig. ٢٢, indicates that the results obtained using different collision pair selection algorithms exhibit excellent agreement with those of the SBT-TAS scheme. All schemes have predicted the same maximum velocity, such as the SB-TAS. Moreover, the results also show the same behavior as the reference solution at the low back pressure region of the cylinder, see Fig. ٢٢-a. The pressure coefficient distribution that is shown in Fig. ٢٢-b is in full agreement with the reference simulation results.

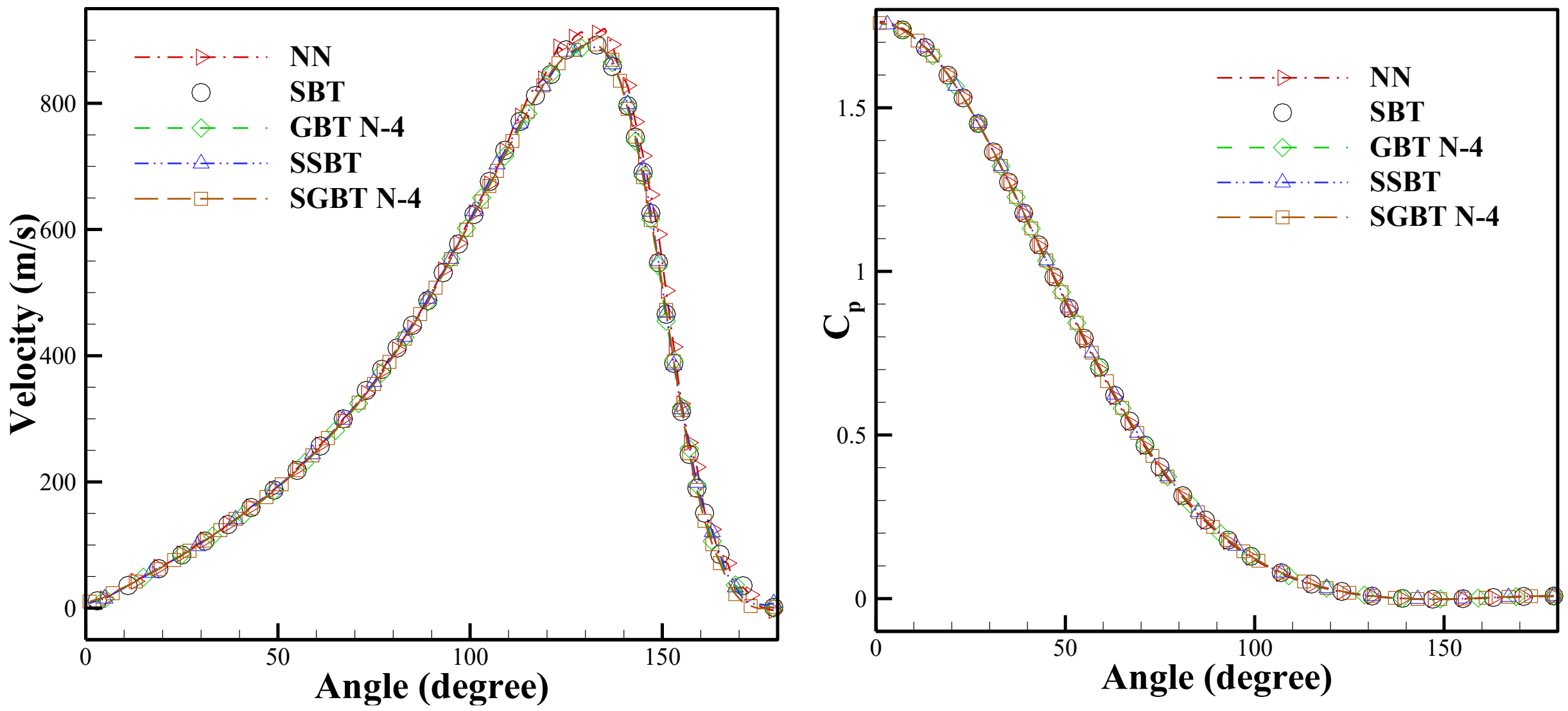

(a) (b)

Fig ۲۲. (a) Velocity magnitude and (b) pressure coefficient distribution on the cylinder surface.

Figure ۲۳-a illustrates the temperature distribution normalized by the free stream temperature along the cylinder surface. The results of the NN, SSBT, GBT, and SGBT schemes with $N_{sel}$ = N-4 are generally in good agreement with the SBT-TAS solution. All schemes predicted the minimum temperature on the cylinder surface at approximately 138 degrees. Some discrepancies are observed near the low-pressure region behind the cylinder. Another investigated parameter is the heat flux on the cylinder surface. It is shown in Figure ۲۳-b. The results of the various schemes are in suitable agreement with the SBT-TAS results. Although the results coincide entirely at the rear region of the cylinder, small deviations are visible in front of the cylinder near the stagnation point region.

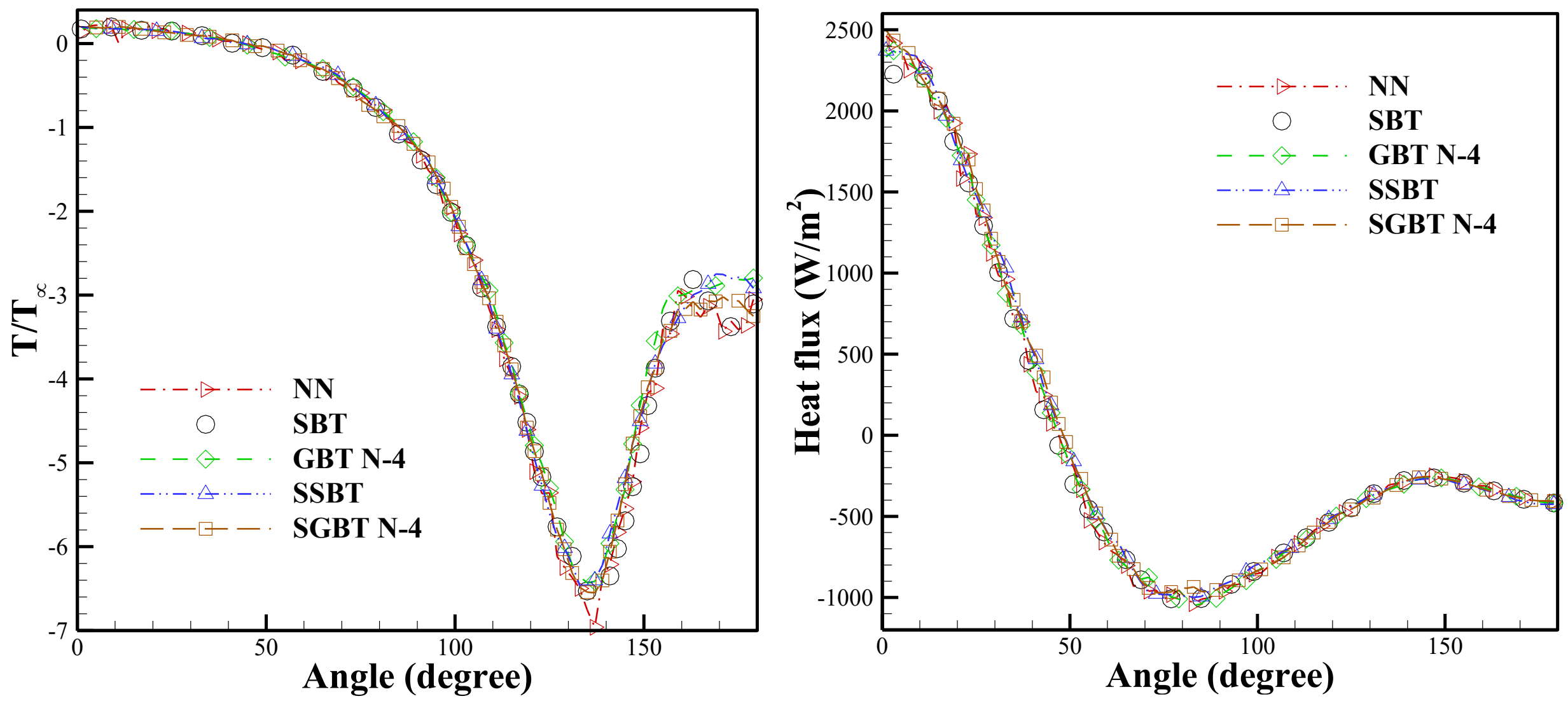

(a) (b)

Fig 23. (a) Normalized temperature and, (b) heat flux on the cylinder surface.

The net heat transfer at the stagnation point is a key parameter analyzed in this study. Table 3 summarizes the results obtained using the SGBT-TAS scheme with various values of $N_{sel}$, ranging from N-2 to N-7, as well as results from other collision partner selection schemes, including SBT, GBT, and SSBT. For comparison, the simulation was also performed using Bird's Nearest Neighbor (NN) scheme, which is the default method implemented in the DS2V code. The NN result is treated as the reference solution, and the relative error of each method concerning this value is reported in the table. From the data, it is evident that the GBT-TAS scheme provides accurate predictions of heat transfer at the stagnation point, with performance comparable to that of other collision selection methods. Among the tested configurations, $N_{sel}$=N-2 and N-3 yield the closest agreement with the NN reference value.

**Table 3.** Heat flux value at the stagnation point.

| $N_{sel}$ | *NN* | *SBT* | *GBT* | *SSBT* | *N-2* | *N-3* | *N-4* | *N-5* | *N-6* | *N-7* |
|---|---|---|---|---|---|---|---|---|---|---|
| $q$ ($W/m^2$) | 2462 | 2327 | 2333 | 2372 | 2445 | 2471 | 2498 | 2366 | 2424 | 2496 |
| *Error (%)* | 0.00% | -5.48% | -5.24% | -3.66% | -0.69% | 0.37% | 1.46% | -3.90% | -1.54% | 1.38% |

Following the methodology employed in our previous work [42] to evaluate the performance of the GBT algorithm in comparison with the SBT and Nearest Neighbor (NN) schemes, a similar performance test is conducted here to assess the computational efficiency of the SSBT and SGBT algorithms across different $N_{sel}$ values. To ensure consistency and enable direct comparison with the results reported in [42], the same simulation conditions are used-specifically, the surface temperature of the cylinder is set to 500 K. In this test, each simulation is run until a predefined convergence criterion is met. The stopping point is defined as the moment when the cumulative normalized difference between the simulated surface heat flux and the benchmark value at all surface points falls below 1. This convergence criterion is expressed as:

$$Error = \sum_{surface\ elements} \frac{Q_{surf} - Q_{benchmark}}{Q_{benchmark}} \leq 1 \tag{25}$$

The reference solution used for this performance analysis is based on the SBT-TAS results, obtained using a sufficiently large number of particles and time-averaged over an extended simulation interval to ensure accuracy and stability. Table 4 presents the results of the performance evaluation for the SSBT and SGBT schemes, compared against the SBT, GBT, and Nearest Neighbor (NN) algorithms. ll simulations were conducted under the same conditions established during the grid and particle independence study. However, for the SSBT and SGBT schemes with $N_{sel}$=N-6, the simulations did not meet the convergence criterion within the standard setup. In these cases, a finer grid was used to facilitate convergence and allow the performance analysis to proceed.

The comparison of CPU time for the SSBT scheme with that of the NN, SBT, and GBT (with $N_{sel}$=N-4) shows that, in this test case, SSBT is slower than the other three methods. As indicated in Table 4, similar to the GBT scheme, the SGBT demonstrates improved computational efficiency over the SBT. However, when compared to the NN method, only the SGBT configuration with $N_{sel}$=N-4, which required the least CPU time to reach the convergence criterion, outperforms NN in terms of speed.

Another key parameter evaluated in this performance analysis is the sample size required to reach the convergence criterion. For clearer comparison, the sample sizes have been normalized concerning those of the NN and SBT schemes. As shown in Table 4, the SSBT algorithm requires a larger sample size than both the NN and SBT to meet the stopping condition, indicating higher computational effort for comparable accuracy. In contrast, the SGBT scheme consistently requires fewer samples than the NN to achieve convergence across various $N_{sel}$ magnitudes. Comparison with SBT, the SGBT scheme required a smaller sample size to reach the stopping point, except for $N_{sel}$=$N$-3 and $N$-6. The higher sample size observed for N-6 can be attributed to the larger initial number of particles used in that simulation. For N-3 is related to the internal algorithm of the SGBT to avoid repeated collisions.

**Table 4.** Results of the performance test analysis, cylinder test case

| | *CPU-time* | *Sample size (×10$^{10}$)* | *Normalized sample size (NN)* | *Normalized sample size (SBT)* | *PPC-grid size* |
|---|---|---|---|---|---|
| *NN* | 13$^{h}$ 46$^{m}$ | 3.0513 | 1 | 2.396 | 20-194×100 |
| *SBT* | 21$^{h}$ 33$^{m}$ | 1.2731 | 0.417 | 1 | 11.5-194×100 |
| *GBT (N-4)* | 9$^{h}$ 49$^{m}$ | 0.7029 | 0.23 | 0.552 | 11.5-194×100 |
| *SSBT* | 35$^{h}$ 45$^{m}$ | 3.4266 | 1.479 | 2.691 | 11.5-250×150 |
| *SGBT (N-3)* | 16$^{h}$ 3$^{m}$ | 1.5513 | 0.508 | 1.218 | 11.5-194×100 |
| *N-4* | 9$^{h}$ 55$^{m}$ | 0.9721 | 0.319 | 0.763 | 11.5-194×100 |
| *N-5* | 18$^{h}$ 15$^{m}$ | 0.8705 | 0.285 | 0.683 | 11.5-194×100 |
| *N-6* | 14$^{h}$ 52$^{m}$ | 1.9517 | 0.639 | 1.533 | 11.5-250×150 |

## 4. Concluding remarks

This study introduces and rigorously evaluates new approaches for collision modeling in the Direct Simulation Monte Carlo (DSMC) method, with a particular emphasis on both theoretical consistency and computational efficiency. The main contributions can be summarized as follows:

1. Markovian NTC-Pre-Scan Algorithm:

   We proposed a modified variant of the classical No-Time-Counter (NTC) method, transforming it into a fully Markovian process through a Pre-Scan procedure. This modification eliminates the non-Markovian memory effects inherent to the traditional NTC, ensuring that the collision process depends only on the current particle ensemble. The Pre-Scan dynamically estimates the local maximum collision product, thereby maintaining accuracy even in challenging conditions with a very low number of particles per cell (i.e., PPC=0.01). The method successfully captures the correct collision frequency across all regimes, including fractional-particle simulations, where conventional NTC methods fail. Note that, even though we improved the NTC scheme by making it Markovian, the problem of repeated collisions in the NTC scheme remains unresolved [44]. This necessitates the need for alternative collision schemes, such as the BT family, which are discussed in detail in the paper. We should note that although the Pre-Scan algorithm introduces a slight computational overhead, this overhead becomes negligible under the low-PPC conditions considered in this paper, as the number of particles per cell is small, and the extra cost can be ignored. This ensures that the method is applied precisely when needed, i.e., in high vacuum conditions when the number of particles is naturally small, and that its computational expense is fully justified under such critical conditions.

2. Systematic Assessment of Bernoulli-Trial (BT) Schemes and Efficiency Gains:

A comprehensive evaluation of simplified (SBT), generalized (GBT), symmetrized (SSBT), and symmetrized–generalized (SGBT) BT schemes was conducted across multiple benchmark problems. These include the Bobylev–Krook–Wu (BKW) relaxation, lid-driven microcavity flow, and hypersonic flow over a cylinder. Results consistently show that BT-based schemes preserve the correct collision frequency and statistical fidelity even when PPC is significantly below unity. Notably, the SGBT scheme demonstrated exact agreement with the fourth-order velocity moment of the Boltzmann solution under the BKW test, validating its physical robustness. Performance analysis revealed significant computational savings. In the cavity problem, the SGBT scheme achieved a reduction of up to ~40% in computational cost with an optimal selection number, while maintaining equivalent or superior accuracy compared to NTC. In the hypersonic cylinder case, the SGBT reached the same level of accuracy with 28% less cost and a sample size of only 0.319 that of the nearest-neighbor method. Such reductions translate directly into practical benefits for large-scale and long-duration DSMC simulations.

A 30–40% reduction in computational cost can make a substantial difference in long-duration simulations or the modeling of large-scale vacuum systems. This demonstrates the practical value of the current work in terms of computational resource savings and aligns directly with the efficiency in industrial vacuum applications. The recommended number of selections for the SGBT scheme is $N_{sel} = N - 3$ or $N - 4$, where $N$ is the number of particles per cell, and was at an order of $N$~20 on the test cases investigated in this work. When the number of particles per cell is relatively large (e.g., $N \approx 20$ or higher), choosing $N_{sel} = N - 3$or $N - 4$ still samples nearly the entire statistical space. Hence, the accuracy is not significantly degraded. In these cases, the acceptance-to-collision ratio remains close to that of the classical NTC algorithm. Although the acceptance rate is not dramatically higher, even this modest reduction in the number of selections is sufficient to yield an overall time savings of approximately 30–40% in simulations involving millions of collisions. At medium to high PPC, the sensitivity to removing a few candidates is very low, so discarding three or four particles has virtually no physical impact while directly reducing the computational burden. Moreover, the weighted exceed ratio (WER) remains below 1% with appropriate choices of $\Delta t$ and $N_{sel}$, ensuring that the physical fidelity of the results is maintained. On the other hand, if $N_{sel}$ is reduced much further, the probability of exceed events (W>1) grows and one would be forced to reduce the time step to maintain stability, which is not practical and would increase the overall computational cost (see for example $N_{sel} = N - 6$ and $N_{sel} = N - 7$ in Table 2). Therefore, the optimal strategy is to decrease the number of selected particles only slightly (e.g., $N - 3$ or $N - 4$), which balances accuracy and efficiency while avoiding unnecessary overhead.

All the algorithms investigated were shown to faithfully reproduce complex physical phenomena, including shock structures in hypersonic rarefied flow and non-equilibrium relaxation dynamics. Advanced statistical diagnostics further confirmed that the schemes preserve fundamental principles of molecular chaos, with correct Poisson-distributed density fluctuations and vanishing spatio-temporal correlations. These advancements have clear relevance for the vacuum community, where DSMC is routinely used in modeling microchannel flows, molecular pumps, and leak detection systems. By enabling accurate simulations with fewer particles and reduced computational expense, the proposed methods broaden the applicability of DSMC to high-vacuum conditions and real-time or large-scale design studies.

In summary, this work provides both a fundamental theoretical correction to the standard NTC model and a practical, high-performance alternative in the form of the SGBT scheme. Together, these contributions not only enhance the reliability and efficiency of current DSMC simulations but also pave the way for next-generation, high-fidelity modeling of complex rarefied gas phenomena. Future work will focus on extending these algorithms to three-dimensional geometries and exploring their hybridization with physics-informed machine learning surrogates, further bridging the gap between algorithmic development and the practical demands of vacuum science and aerospace engineering.

**Acknowledgement**

Stefan Stefanov was partially supported by the Centre of Excellence in Informatics and ICT established under Grant No. BG05M2OP001-1.001-0003, financed by the Science and Education for Smart Growth Operational Program and co-financed by the European Union through the European Structural and Investment Funds.